\documentclass[aps,prb, 
superscriptaddress,
reprint,
longbibliography,
floatfix]{revtex4-1}

\usepackage{textcomp}
\usepackage{amsmath}
\usepackage{amssymb}
\usepackage{array,multirow}
\usepackage{mathtools}

\usepackage{graphicx}
\usepackage{float}
\usepackage{dcolumn}
\usepackage{bm}
\usepackage{caption}
\usepackage{subcaption}
\usepackage[english]{babel}
\usepackage[T1]{fontenc}
\usepackage[utf8]{inputenc}
\usepackage{mathrsfs}
\usepackage{hyperref}

\newcommand{\bra}[1]{{\langle #1 |}}
\newcommand{\ket}[1]{{| #1 \rangle}}

\newcommand{\kv}{{\bf k}}

\newcommand{\ep}{\epsilon}

\newcommand{\bsig}{{\bm{\sigma}}}

\newcommand{\alp}{\alpha}
\newcommand{\bb}{\beta}

\newcommand{\xh}{\hat{x}}
\newcommand{\zh}{\hat{z}}

\DeclareUnicodeCharacter{2212}{-}

\begin{document}


\title{Electron transport through antiferromagnetic spin textures and skyrmions in a magnetic tunnel junction}

\author{Nima Djavid}
\email{ndjav001@ucr.edu}
\author{Roger K. Lake}%
\email{rlake@ece.ucr.edu}
\affiliation{%
Department of Electrical and Computer engineering, University of California Riverside,
Riverside 92521, USA
}%

\date{\today}

\begin{abstract}
An ideal layered $\xh$-polarized antiferromagnet (AFM) between two antialigned 
$\pm \zh$ polarized ferromagnetic (FM) contacts  
transmits no current due to a $\pi$ phase difference of the matrix elements
coupling the spin degenerate states to the two FM contacts.
%
%
Inserting a normal metal layer or tunnel barrier layer between one FM contact
and the AFM alters this phase difference, 
and, due to the unequal weighting of the two spins
at the interface, it also breaks the spin degeneracy of the two 
AFM states. 
The broken symmetry of the matrix elements combined with the broken degeneracy of the AFM states,
result in a Fano resonance in the transmission and a turn-on of the $T_{\uparrow,\downarrow}$
transmission channel.
Such a magnetic tunnel junction geometry with two antialigned $\pm \zh$ FM 
contacts can electrically detect an AFM skyrmion. 
The AFM skyrmion serves as an analogue of the oblique polarizer in the triple polarizer experiment.
Resistances and resistance ratios are calculated and compared for FM and AFM skyrmions in a magnetic tunnel junction.
\end{abstract}

\maketitle


\section{\label{sec:level1}Introduction}
There is rapidly growing interest in  
antiferromagnetic (AFM) materials for use as the active elements of spintronic devices
\cite{AFM_spintronics_Jungwirth_NNano16,2017_AFM_spintronics_Jungwirth_PSSR,2018_Tserkovnyak_RMP}.
Their lack of macroscopic magnetic fields allows AFM devices and interconnects to be 
highly scaled with reduced cross talk and insensitivity to geometrical anisotropy effects.
AFMs have resonant frequencies, magnon velocities, and switcing speeds that 
are several orders of magnitude higher than those
in ferromagnetic (FM) materials
\cite{AFM_spintronics_Jungwirth_NNano16,KWang_ULowSwitchingAFM_APL16}.
The opportunities of speed, scaling, and robustness to stray fields come with challenges. 
The insensitivity to external fields makes both the manipulation and detection of the AFM order parameter difficult.
\begin{figure}[t]
\includegraphics[width=3.4in]{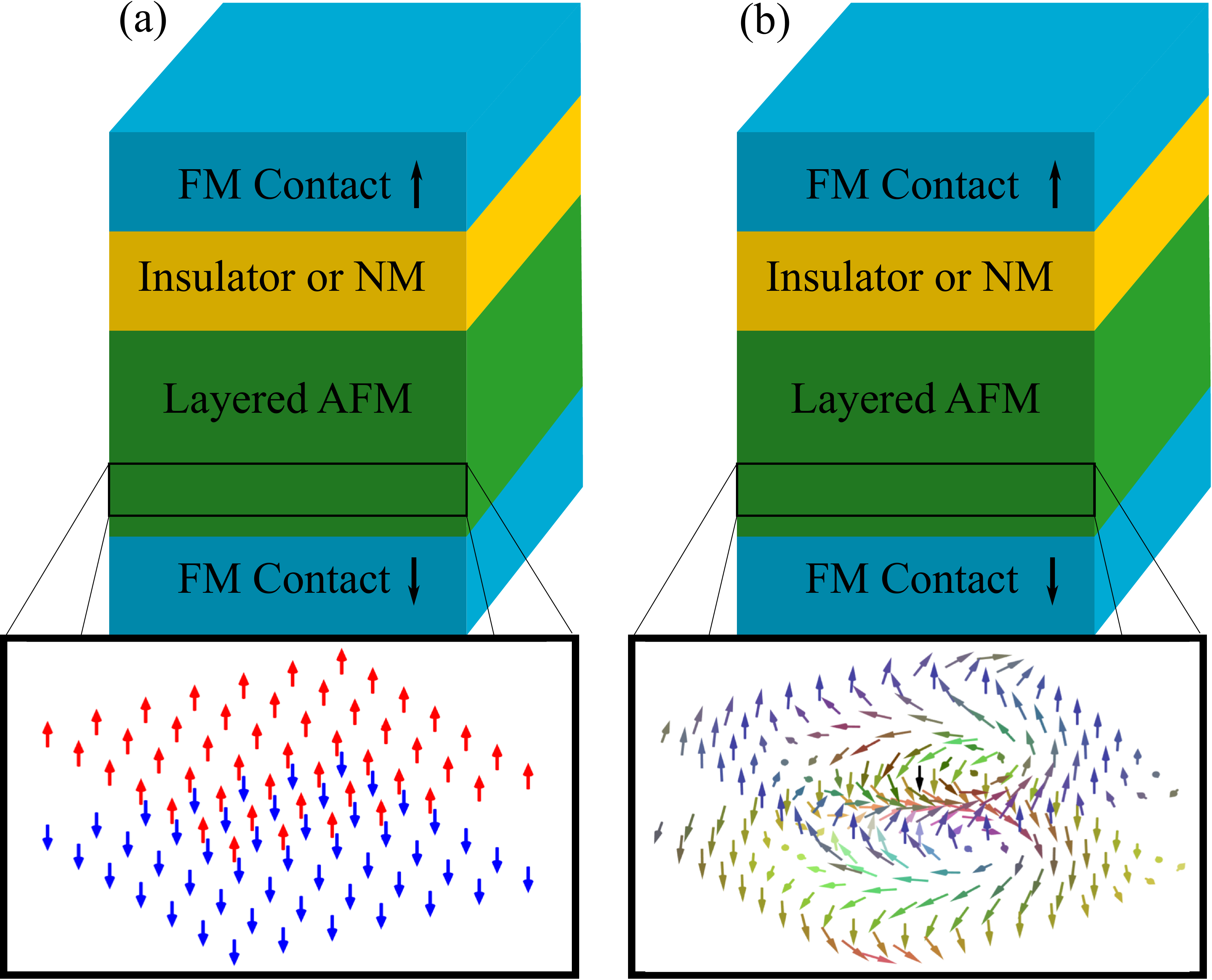} 
\caption{
A magnetic tunnel junction consisting of 
ferromagnet/ insulator/ anti-ferromagnet/ ferromagnet layers.
The AFM layer consists of a compensated, aligned, layered AFM with a N{\'e}el
vector that is either
(a) out-of-plane or (b) in an AFM skyrmion texture.
The top and bottom FM contacts have 
perpendicular magnetic polarization, and their polarizations are anti-parallel
corresponding to the high resistance state of a traditional MTJ.
The spacing layer between the top contact and AFM can be a tunnel barrier or normal metal (NM).
}
\label{fig:AFMSchematic}
\end{figure}

In this work, we consider the magnetoresistance of a magnetic tunnel junction (MTJ)
with a layered AFM metal inserted between the barrier and lower FM contact as shown in Fig. \ref{fig:AFMSchematic}.
The polarization of the two FM contacts are anti-parallel with the magnetic moments in the 
$\pm \zh$ directions. 
This is the configuration of the high resistance state of a MTJ.
However, if a FM layer with magnetization oriented in the $\xh$ direction is placed between the two FM contacts,
it serves as an analogue of the oblique polarizer \cite{Dirac_PQM} in the three-polarizer experiment, 
and it opens up the $T_{\uparrow,\downarrow}$
transmission channel reducing the resistance to within a factor of $1/2$ of the aligned value.
We now ask the question, ``Will a collinear AFM layer with its N\'{e}el vector oriented in the $\xh$ direction
also act as an analogous oblique polarizer in the MTJ?''
In a collinear AFM, each spin ${\bf S}$ is paired with it opposite spin $-{\bf S}$ in an AFM unit cell. 
Since the sum and difference of the eigenstates of $s_x$ give the eigenstates of $s_z$, it is not
clear that a $\xh$ polarized AFM layer will act in the same way as a $\xh$ FM layer.

However, it is clear that the orientation of an AFM layer can affect the tunneling magnetoresistance 
(TMR).
Theoretical work analyzing the magnetoresistance of misaligned, collinear AFM layers does find that
N\'{e}el vector alignment affects the 
resistance. \cite{MacDonald_AFM_Spin_Torques_PRB06,
Cr_Au_Cr_HGuo_MacDonald_PRB07,
STT_AFM_AR_FeMn_DFT_Xia_PRL08,
Fe_MgO_FeMn_Cu_TAMR_PRB17,
Mn3Pt_SrTiO3_Pt_TMR_PRAppl19,
Mn2Au_MTJ_ScChina20}
Experimentally, resistance changes due to N\'{e}el vector alignment are observed in the 
tunneling anisotropic magnetoresistance.\cite{TAMR_AFM_Jungwirth_2011,IrMn_TMR_Pan_PRL12,MnPt_Piezo_NNano19}

Due to magnetic crystalline anisotropy, the  N\'{e}el vector of the AFM layer will prefer to align
along a given axis.
If we assume perpendicular magnetic anisotropy, the  N\'{e}el vector will align 
in the $\pm \zh$ direction. 
To rotate and stabilize the N\'{e}el vector in the $\xh$ direction requires altering
the magnetic anisotropy such as by applying strain \cite{MnPt_Piezo_NNano19,InJunPark_APL19}.
Another way to obtain local $\xh$ components of the N\'{e}el vector is by the presence of a local
topological spin texture such as a skyrmion.
There have been a number of theoretical investigations of creation, stability,
and control of AFM and synthetic-AFM 
skyrmions \cite{2016_J_ind_AFM-Sky_motion_NJP,
AFM_driven_by_SHE_APL16,
2016_AFM_Skx_Stab_Creat_Manip_Ezawa_SciRep,
2016_AFM-Skx_J_T_Tretiakov_PRL,
2017_AFM_Skx_crystals_PRB,
THE_AFM_J_induced_motion_Tretiakov_PRL18,
2018_Skx_injection_NJP,
AFM_Skx_FET_APL18,
Stability_lifetime_Tretiakov_PRB19,
2019_UFast_Gen_Dyn_Brataas_PRB}.
Recent experimental work shows that ultra-small size AFM skyrmions can 
be stabilized at room temperature in synthetic antiferromagnets
and Heusler compounds \cite{RT_Stabilization_ASKn_Fert19,Felser_Parkin_ASk_Heusler_NCom19}. 
Theoretical estimates predict that skyrmion diameters can reduced
below 20 nm in synthetic AFMs \cite{RT_Stabilization_ASKn_Fert19}.
%
%

Methods for detecting AFM skyrmions include topological spin hall measurements \cite{Buhl2017} 
and magnetic force microscopy \cite{Legrand2019}. 
However neither of the above approaches are well suited for application in highly scaled 
memory devices.
From the perspective of geometry and scaling, the MTJ is an ideal structure due to its
minimal in-plane cross-sectional area and two-terminal operation.\cite{MRAM_Review_TED20}
For FM skyrmions, electrical detection has been heavily investigated \cite{Hanneken2015a, Kubetzka2017, Maccariello2018a, Wang2019, Hamamoto2016, Tomasello2017, Stolt2019, MTJ_Skx_Detection_PRB19}.
Among these studies, electrical detection of single FM skyrmions in an MTJ geometry was theoretically investigated in
Ref. \cite{MTJ_Skx_Detection_PRB19}.  
We will show that the presence or absence of an AFM skyrmion in the MTJ structure of Fig. \ref{fig:AFMSchematic}(b) 
also gives rise to a magnetoresistance
difference suitable for electrical detection. 

The paper is organized as follows.
Sec. \ref{sec:methods} describes the methods used for the numerical calculations. 
The N\'{e}el vector dependence of the magnetoresistance of the structure shown in Fig. \ref{fig:AFMSchematic}(a)
is analyzed in Sec. \ref{sec:1DChain}.
The magnetoresistances due to the presence of both FM and AFM skyrmions are described in Sec. \ref{sec:3D_TMR}.
Summary and conclusions are given in Sec. \ref{sec:conclusions}.
\begin{figure*}[t]
     \begin{subfigure}[b]{0.32\textwidth}
         \includegraphics[width=\textwidth]{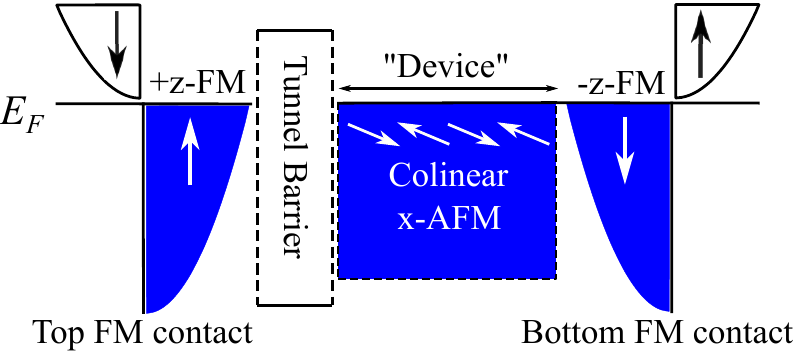} 
         \caption{}
     \end{subfigure}
     \begin{subfigure}[b]{0.32\textwidth}
         \includegraphics[width=\textwidth]{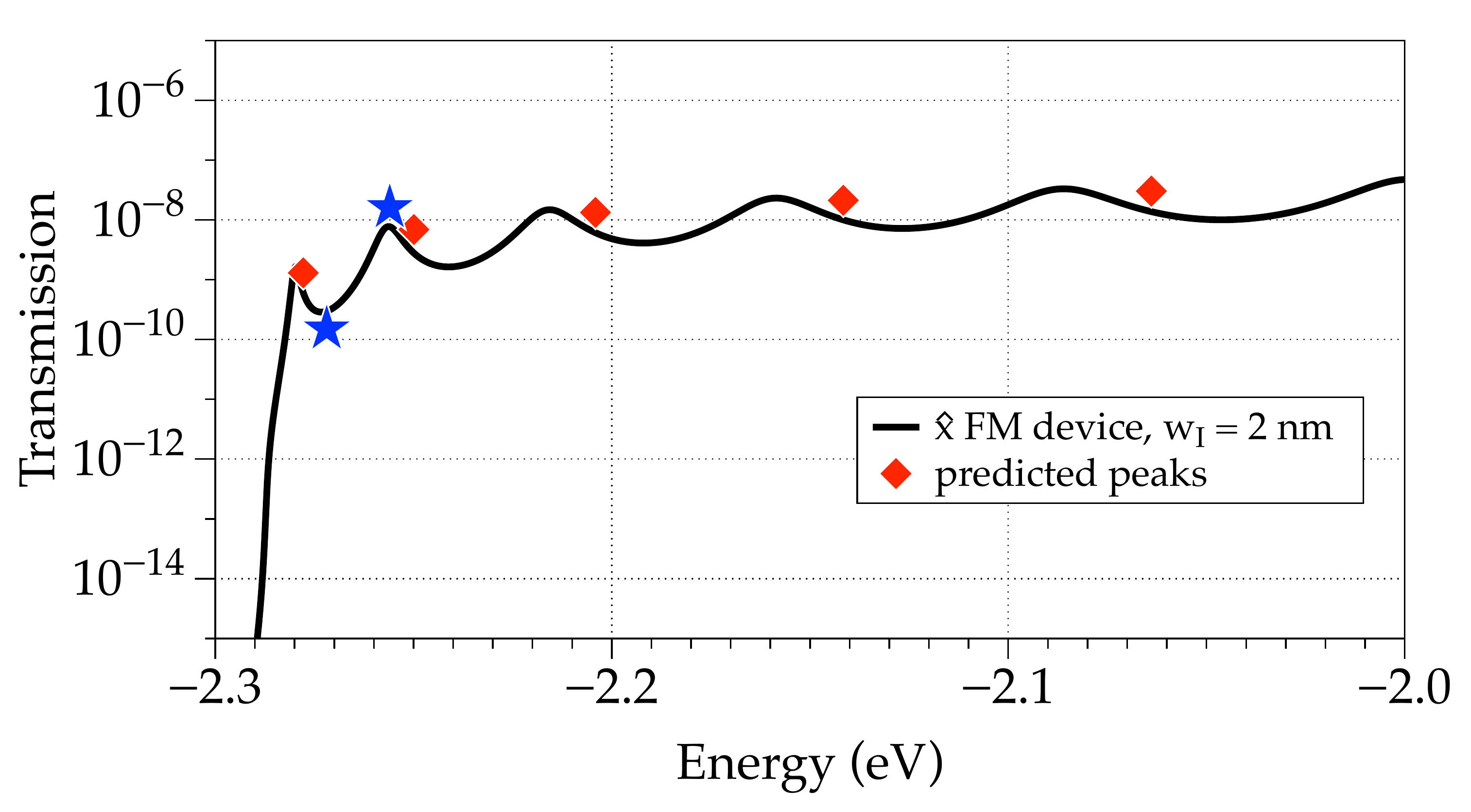} 
         \caption{}
     \end{subfigure}
     \begin{subfigure}[b]{0.32\textwidth}
         \includegraphics[width=\textwidth]{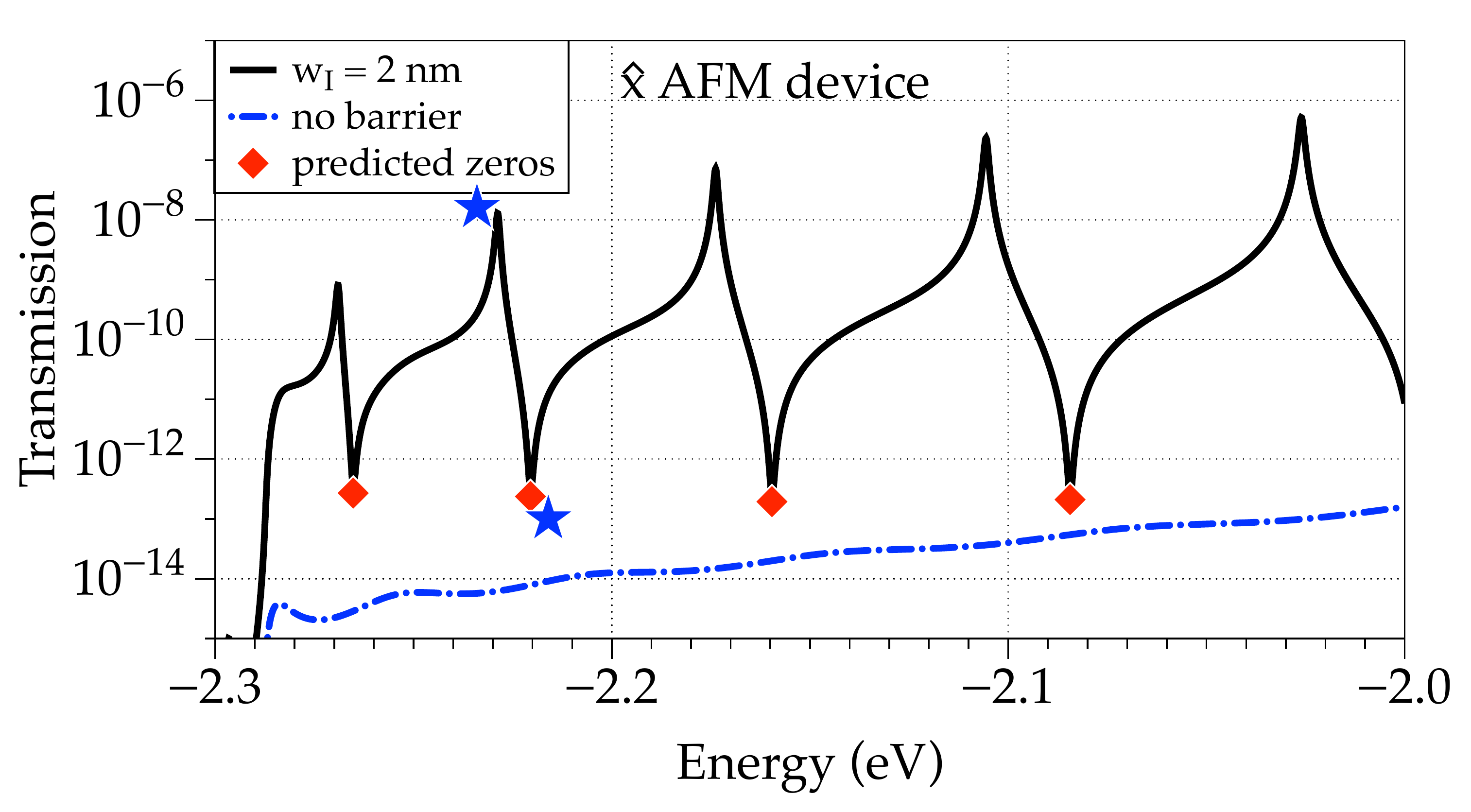} 
         \caption{}
     \end{subfigure}
     \vspace{2em}
     \begin{subfigure}[b]{0.32\textwidth}
         \includegraphics[width=\textwidth]{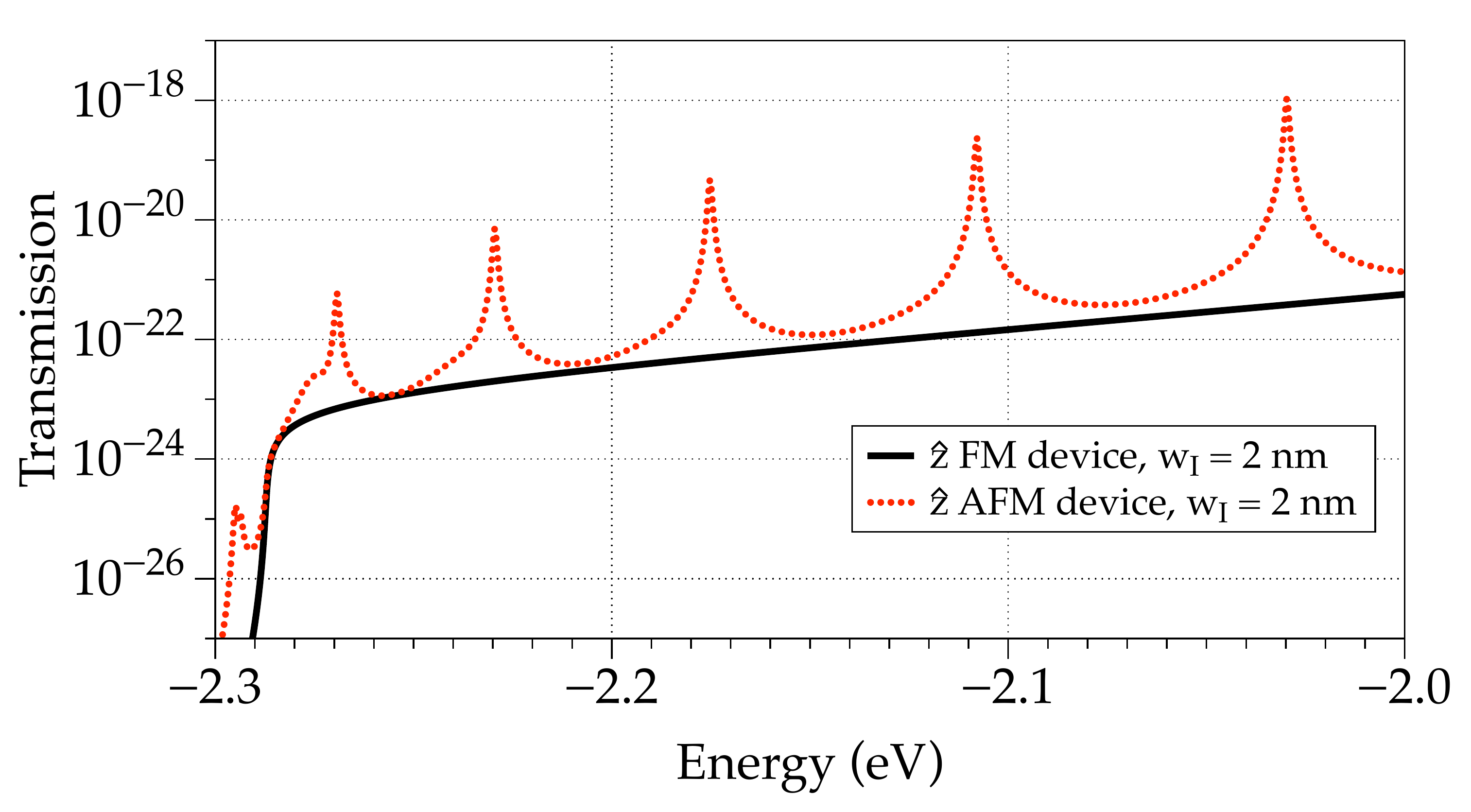} 
         \caption{}
     \end{subfigure}
     \begin{subfigure}[b]{0.32\textwidth}
         \includegraphics[width=\textwidth]{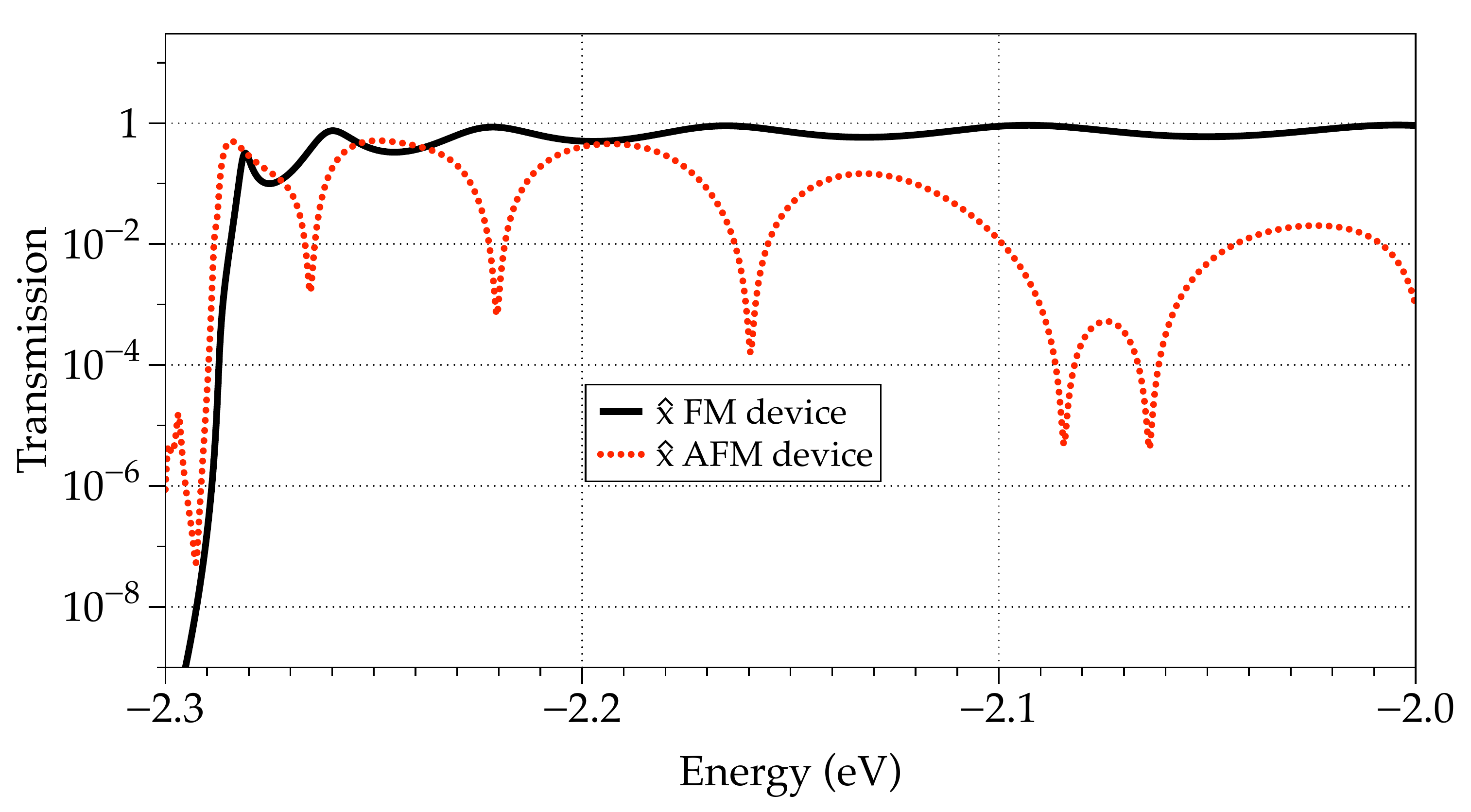} 
         \caption{}
     \end{subfigure}
     \begin{subfigure}[b]{0.32\textwidth}
         \includegraphics[width=\textwidth]{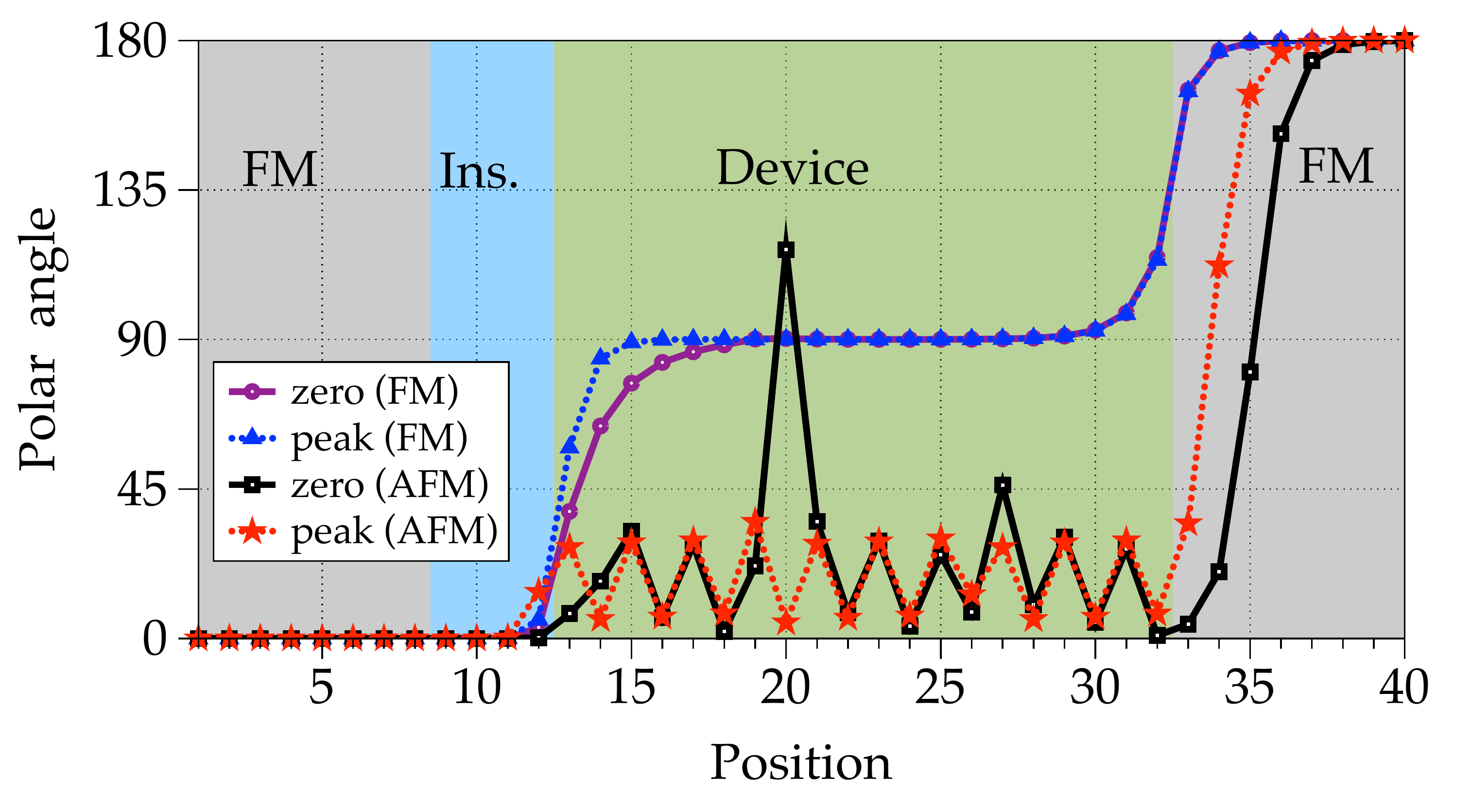} 
         \caption{}
     \end{subfigure}
\caption{
(a) Schematic showing the 4 different regions: up-spin polarized top contact, tunnel barrier, 
FM or AFM region with various spin alignments, and down-spin polarized bottom contact.
(b) - (e) $T_{\uparrow , \downarrow}$ transmission with the device region in a 
(b) x-polarized FM state and a
(c) x-polarized AFM state.
The lower dashed-dot blue curve in (c) shows the same transmission when the tunnel barrier is removed.
(d) $T_{\uparrow , \downarrow}$ transmissions with the device region is in a
z-polarized FM or AFM state. 
(e) Transmission of the x-polarized FM and AFM devices when the tunnel barrier is
replaced with a normal metal layer.
(f) 
Expectation value of the polar angle of the spin at each lattice site for the 
x-FM and x-AFM device. 
For the FM device, the two angles are shown for the peak and valley indicated
by the astericks in (b), and for the AFM device, the
two angles are shown for the peak and the zero designated by the astericks in (c).
}
\label{fig:1D_results}
\end{figure*}

\section{Methods}
\label{sec:methods}
The model tight-binding Hamiltonian for the system of Fig. \ref{fig:AFMSchematic} 
is \cite{Nagaosa_Hunds_Rule_Coupling_PRL00,MacDonald_AFM_Spin_Torques_PRB06}
\begin{equation}
H = \sum_{i} {\bf c}_{i}^\dagger \ep_i {\bf c}_{i} 
+ \sum_{<i,j>}({\bf c}_i^\dagger t_0 {\bf c}_j + H.c.) - J \sum_i {\bf c}_i^\dagger 
{\bm{\sigma}}_i \cdot {\bf S}_i {\bf c}_i,
\label{eq:Hgeneric}
\end{equation}
where $i,j$ are the site indices, and $\sum_{\langle i,j \rangle}$ 
indicates a sum over all nearest neighbors.
${\bf c}_i = [ c_{i,\uparrow} \; c_{i,\downarrow} ]^T$ is the spinor
annihilation operator for site $i$.
${\bf S}_i$ is localized spin on site $i$, 
$\bsig_i$ is the electron spin on site $i$,
and $J$ is the Hund's coupling.
$t_0$ is the nearest neighbor hopping term, 
and $\ep_i$ is an on-site energy term used to create the insulating barrier and align the bands
in the different regions.
In the insulating region, $\ep_i = \pm 3.8$ eV 
alternates between layers in the $z$ direction mimicking alternating layers of anions and cations.
This results in an insulating gap at $\Gamma$ of $7.8$ eV, which is the same as MgO.
In the AFM and FM regions, $\ep_i$ is constant and only serves to align the bands of the different regions.
With this model, the tight binding dispersions for the AFM, insulating, and FM regions are
\begin{align}
E_{\rm AFM}&= \pm \sqrt{J^2 + 2t_0^2(1+\cos(k_za))} + \ep_t(k_x,k_y)
\label{eq:AFM_Ham} \\
E_{\rm I}&= \pm \sqrt{\ep_i^2 + 2t_0^2(1+\cos(k_za))} + \ep_t(k_x,k_y)
\label{eq:INS_Ham} \\
E_{\rm FM}&= \pm J + 2t_0\cos(k_za) + \ep_t(k_x,k_y)
\label{eq:FM_Ham}
\end{align}
where $\ep_t(k_x,k_y) = 2t_0\cos(k_xa) + 2t_0\cos(k_ya)$.
With $t_0 = -0.3812$ eV and $a = 5$ \AA, m* = 0.4$m_e$ which corresponds to the effective mass of MgO. 
%
%

%
%
%
%

%

The tunneling magnetoresistance is calculated from the conductance given by 
\begin{equation}
G = -\frac{e^2}{h} \sum_{\sigma , \sigma'} \int \ \frac{d^2k}{4\pi^2}  \int dE \: T_{\sigma, \sigma'}(E,\kv)  
\left( \frac{\partial f(E-E_F)}{\partial E} \right)
\label{eq:G}
\end{equation}
where the transmission coefficent is  
\begin{equation}
T_{\sigma, \sigma'} (E) = {\rm trace} \{ \Gamma_{\sigma, \sigma}^{\mathcal L} G^R_{\sigma, \sigma'} 
\Gamma_{\sigma', \sigma'}^{\mathcal R} G^A_{\sigma',\sigma} \} ,
\label{eq:T}
\end{equation}
and $\kv$ is the two dimensional wavevector in the $x-y$ plane.
In Eqs. (\ref{eq:T}), the retarded Green's function is
\begin{equation}
G^R(E,\kv) = [E - H(\kv) - \Sigma_L(E,\kv) - \Sigma_R(E,\kv)]^{-1},
\end{equation}
where $\Sigma_{L,R}$ are the self energies due to the semi-infinite FM contacts,
$G^A = {G^R}^\dagger$, and
$\Gamma_{L,R} = - 2 {\rm Im} \{\Sigma_{L,R} \}$.
Numerically, the contact self energies are calculated using the decimation method \cite{MPLSancho_HighlyConvergent_JPF85}
with a an imaginary term $i\eta$ on the diagonal elements with $\eta = 0.5$ meV.
In the analysis of Sec. \ref{sec:1DChain}, an exact analytical expression is also used.
For the 3D devices in Sec. \ref{sec:3D_TMR}, 
the conductance is calculated at zero temperature so that $-\partial f/\partial E = \delta(E-E_F)$,
and the Fermi level is chosen such that the top contact is 100\% polarized. 
Thus, Eq. (\ref{eq:G}) becomes
\begin{equation}
G = -\frac{e^2}{h} \sum_{\sigma} \int \ \frac{d^2k}{4\pi^2} \: T_{\uparrow , \sigma}(E_F,\kv) . 
\label{eq:G_analy_0_T}
\end{equation}

To simulate MTJ structures with skyrmions, 
a $15 \times 15$ site supercell is used with periodic boundary conditions
in the $x$-$y$ plane. 
The integral over $k_x$ and $k_y$ is performed using noramalized $k'_{x,y} \in [0, \pi]$. The momentum discretization length is chosen to be $dk'_x = dk'_y = \pi/31$.
The resistance is $R = A_{cs}/G$, which $A_{cs}$ is the area of the supercell in the $x$-$y$ plane.

For the 1D spin chain, the expectation value of the spin at each site is calculated
from the left injected spinor wavefunction, normalized at each site, and given by
\begin{equation}
{\boldmath \psi}^{\mathcal L}_n
=
\left[
\begin{array}{c}
G^R_{1 \uparrow,n\uparrow} \\ 
G^R_{1 \uparrow,n\downarrow} 
\end{array}
\right]
/ \sqrt{ \left| G^R_{1\uparrow,n\uparrow} \right|^2
+ \left| G^R_{1\uparrow,n\downarrow} \right|^2 }.
\end{equation}
The expectation value of the polar angle at each site is then 
\begin{equation}
\theta_n = \arccos ( \bra{ \psi^{\mathcal L}_n} \sigma_z \ket{\psi^{\mathcal L}_n} ).
\label{eq:theta_n}
\end{equation}

The magnetization of a FM skyrmion or the N\'{e}el vector of a single skyrmion is described by
\begin{equation}
{\bf n}(r) = [\sin \gamma (r) \cos \varphi (\phi), \sin \gamma (r) \sin \varphi (\phi), \cos  \gamma (r)]
\end{equation}
where $\varphi(\phi) = m\phi + \upsilon$, $\phi = \tan^{-1}(\frac{y}{x})$, 
and $\upsilon$ determines the helicity of skyrmion.
Since a layered type AFM is considered,
the AFM skyrmions are constructed as FM skyrmions in each layer with ${\bf n}$ reversed in alternating
layers.
We consider both Bloch-type (m = 1, $\upsilon$ = $\frac{\pi}{2}$)
and N\'{e}el type (m = 1, $\upsilon$ = $0$) skyrmions with
$\gamma(r=0) = 0$, $\gamma(0<r<R) = \pi(1-\frac{r}{R})$, and $\gamma(r>R) = \pi$,
where $R$ is the radius of the skyrmion.
The diameter of the skyrmion is 13 gridpoints. 
Since the discretization length is $5$ \AA, the skyrmion diameter is $6.5$ nm. 
Although the size of skyrmion is small, it includes enough sites to capture the physics we discuss here.
%

\section{Transmission through a layered AFM in a MTJ}
\label{sec:1DChain}
%
The layered structure that we first consider is schematically illustrated in Fig. \ref{fig:1D_results}(a).
It consists of two 100\% polarized anti-aligned FM contacts with the spin polarized
in the $\pm \hat{z}$ direction.
In the central region, we consider both FMs and layered AFMs polarized in the $\pm \hat{z}$
or $\pm \hat{x}$ directions.
We refer to this region as the ``device region.''
For the calculations shown in Fig. \ref{fig:1D_results}, 
its thickness is fixed at $10$ nm (20 layers).
This is for illustration purposes, since it allows multiple resonant peaks to be shown.
The system is considered both with and without the tunnel barrier.
The insulating barrier thickness ($W_I$) is $2$ nm.
The Hund's coupling in the contacts is $J = 500$ meV to enforce 100\% polarization,
and in the device it is 120 meV. 
The nearest neighbor hopping term is $t_0 = -0.38$ eV. 
We begin by presenting numerical $\Gamma$-point calculations of the transmission 
for various spin alignments in Fig. \ref{fig:1D_results},
and then we provide an analytical analysis of transmission through the AFM layer.

When the device region is either FM or AFM and polarized in the $\pm \hat{z}$ direction,
the $T_{\uparrow , \downarrow}$ transmission is suppressed by more than 18 orders 
of magnitude
as shown in Fig. \ref{fig:1D_results}(d).
The conductance resulting from this channel would be unmeasurable.  

When the device region is FM polarized in the $\hat{x}$ direction, i.e. $S_i = \hat{x}$
in the device region, the transmission of the spin flip channel increases
over 10 orders of magnitude
to approximately $10^{-8}$ as shown in Fig. \ref{fig:1D_results}(b). 
The $x$-polarized region serves as the oblique polarizer that allows transmission
between the anti-aligned contacts. 
The red diamonds correspond to the particle-in-a-box energies corresponding to 
integer half-wavelengths confined in the device region 
between the tunnel barrier and the bottom FM contact.
Using the tight binding dispersion, Eq. (\ref{eq:FM_Ham}), these energies are
$E_{\rm FM}(n) = - J_H - 4|t_0| - 2|t_0| \cos \left( \tfrac{n \pi}{N+1} \right)$ 
where $n = 1, \: 2, \: 3, \cdots$,
and $N = 20$ is the number of lattice sites in the FM device region.
The lowest resonances in the transmission are close to the energies of 
a confined wavefunction. 
As the resonance energies increase, the resonance peaks fall further below the 
energy of a confined state, since the higher energy states are less well confined
by a finite confining potential.
While the tunnel barrier creates good confinement on the left, the matrix
element to the right $-\hat{z}$ polarized contact is only reduced by a factor
of $1/\sqrt{2}$ compared to that when the spins of the device and contact are aligned.
Therefore, the resonances are very broad as expected for weak confinement.
For the $\hat{x}$ polarized FM device, 
the tunnel barrier only serves to reduce the overall conductance.
Without the tunnel barrier, the resonant transmission peaks reach $1.0$,
i.e. perfect transmission,
as expected for equal coupling to the two contacts.

Fig. \ref{fig:1D_results}(f) shows the expectation value of the polar angle of the spin
at each site calculated from Eq. (\ref{eq:theta_n}) for wavefunctions
calculated at the energies corresponding to the two blue stars in 
Figs. \ref{fig:1D_results}(b) and \ref{fig:1D_results}(c).
Once the $+\hat{z}$ electron tunnels through the barrier, it becomes polarized in the 
$\hat{x}$ direction in the FM device. 
This $\hat{x}$ polarized electron can then pass into the $-\hat{z}$ polarized
right contact in complete analogy with the three polarizer experiment.

When the $\hat{x}$ polarized FM device is replaced with an AFM device with an
$\hat{x}$ polarized N\'{e}el vector, the picture becomes more interesting.
The transmission of the system is shown in Fig. \ref{fig:1D_results}(c).
The broad resonances resulting from the $x$-AFM device 
of Fig. \ref{fig:1D_results}(b) are replaced by sharp Fano resonances.
A Fano resonance results when there are two transmission paths
in which an extended state is weakly coupled to a localized state \cite{Ugo_Fano_PRL61}. 
In the $\xh$-AFM device, the two transmission paths are provided by the 
doubly degenerate AFM band.
The red diamonds in Fig. \ref{fig:1D_results}(c) show the energies corresponding to
integer half wavelengths in the device region of the AFM band.
For the bipartite AFM lattice, these energies are
given by Eq. (\ref{eq:Ep}).
%
%
%
In a simple model of a bound state weakly coupled to a 1D chain, the zeroes of the transmission
occur at exactly the unperturbed energy of the bound state.
In this case, the zeroes of the transmission occur at the energies of ideal particle-in-a-box
states bound in the $\xh$-AFM device region.

The expectation value of the polar angles calculated at the peak and zero indicated by the blue stars 
in Fig. \ref{fig:1D_results}(c) are shown in Fig. \ref{fig:1D_results}(f) as a function of position.
At the transmission peak, the spin starts to pick up an in-plane component while still in the
tunnel barrier, and maintains a finite in-plane component at every site within
the $\xh$-AFM region. 
At the transmission zero, the spin remains $\zh$ polarized throughout the barrier, only acquires
an in-plane component at the first site of the $\xh$-AFM region, and is 100\% $\zh$ spin polarized at
the last site of the $\xh$-AFM region.

A more surprising result from the $\xh$-AFM device is that when the tunnel barrier is removed,
the transmission falls orders of magnitude as shown by the blue curve
in Fig. \ref{fig:1D_results}(c).
This is contrary to expections, since one would expect that in the absence
of a tunnel barrier, the conductance should exponentially increase,
such as it does for the $\xh$-FM device.
Thus, the non-magnetic tunneling region must play a critical role in 
coupling the two degenerate AFM bands to the contacts and to each other.
If we replace the tunneling region with a normal metal region, then we find that the 
Fano resonances remain, but the transmission uniformly increases, 
such that the transmission peaks approach 1.0.

To understand the role of a tunneling region or a normal metal region, we 
consider a $\xh$-AFM region sandwhiched between two anti-aligned $\zh$-FM contacts
with no tunnel barrier.
We work in the orbital basis of the eigenstates of the finite $\xh$-AFM device and 
in the spinor basis of the eigenstates of $\sigma_x$, which we will refer to
as the $\{ \ket{X} \}$ basis.
The FM contact regions are included through self-energies on the first and last sites of the
$\xh$-AFM device.

For a periodic $\xh$-AFM system, the unit cell consists of 2 lattice sites labeled as the 
$\alp$ site and the $\bb$ site.
The lattice constant corresponding to the two sites is $a_{\rm uc}$.
The spinor  $\{ \ket{X} \}$ basis in which we represent the Hamiltonian is
 $\{ \bm{\chi}_\alp^+, \bm{\chi}_\alp^-, \bm{\chi}_\bb^+, \bm{\chi}_\bb^-  \} = 
\{ \tfrac{1}{\sqrt{2}}[1 \; 1 \: 0 \: 0]^T, \tfrac{1}{\sqrt{2}}[1 \; \text{-}1 \: 0 \: 0]^T, \tfrac{1}{\sqrt{2}} [0 \; 0 \: 1 \: 1]^T,  \tfrac{1}{\sqrt{2}} [0 \; 0 \: 1 \: \text{-}1]^T \}$
where the order corresponds to $[\alp \uparrow, \alp \downarrow, \bb \uparrow, \bb \downarrow ]$.
In this basis, the Hamiltonian matrix resulting from Eq. (\ref{eq:Hgeneric}) with ${\bf S}_i = \pm \xh$ is 
\begin{equation}
H = 
\left[
\begin{array}{cccc}
-J & 0 & t_k & 0 \\
0 & J & 0 & t_k \\
t_k^* & 0 & J & 0 \\
0 & t_k^* & 0 & -J
\end{array}
\right] .
\end{equation}
where $t_k = t_0(1+e^{-ika_{\rm uc}}) = 2t_0\cos(ka_{\rm uc}/2) e^{-ika_{\rm uc}/2}$.
The 2 degenerate band eigenenergies are $E = \pm \ep_k$
where $\ep_k = \sqrt{J^2 + 4t^2\cos^2(\tfrac{ka_{\rm uc}}{2})}$.
The two eigenvectors of the degenerate lower band, normalized within each unit cell, are
\begin{align}
\ket{\zeta^-_{+x}} &= \left[ \sqrt{ 1 + \tfrac{J}{\ep_k} } \ket{ \chi_\alp^+ } 
+ 
e^{ika_{\rm uc}/2} \sqrt{ 1 - \tfrac{J}{\ep_k} } \ket{ \chi_\bb^+ } \right] \tfrac{e^{ikn_{\rm uc}a_{\rm uc}}}{\sqrt{2}} \\
\ket{\zeta^-_{-x}} &= \left[ \sqrt{ 1 - \tfrac{J}{\ep_k} } \ket{ \chi_\alp^- } 
+ 
e^{ika_{\rm uc}/2} \sqrt{ 1+ \tfrac{J}{\ep_k} } \ket{ \chi_\bb^- } \right] \tfrac{e^{ikn_{\rm uc}a_{\rm uc}}}{\sqrt{2}}
\end{align}
where $n_{\rm uc}$ is the index of the unit cell.

In the finite length uncoupled device, the plane wave solutions become the standing wave solutions
with discrete wavevectors $k_p = \frac{p\pi}{(N_{\rm uc} + \frac{1}{2} ) a_{\rm uc} }$ 
where $p \in \{1,2,\cdots,N_{\rm uc} \}$ and $N_{\rm uc}$ is the number of unit cells in the AFM
region. 
The discrete eigenenergies are
\begin{equation}
E_{\rm AFM}(p) = \pm \sqrt{J^2 + 4t_0^2 \cos^2 \left( \tfrac{p \pi}{2N_{\rm uc} + 1} \right)} .
\label{eq:Ep}
\end{equation}
The two degenerate eigenstates for each discrete energy within the lower band are
\begin{align}
\ket{\xi_{1},p} =& 
\tfrac{1}{\sqrt{N_{\rm uc}+1/2}}
\left[
\sin \left( \tfrac{(n_{\rm uc} - \frac{1}{2}) \pi p }{N_{\rm uc} + \frac{1}{2}} \right)
\sqrt{ 1 + \tfrac{J}{\ep_p} } \ket{ \chi_\alp^+ } \right.
\nonumber \\
&+
\left.
\sin \left( \tfrac{n_{\rm uc} \pi p }{N_{\rm uc} + \frac{1}{2}} \right)
\sqrt{ 1 - \tfrac{J}{\ep_p} } \ket{ \chi_\bb^+ } 
\right]
\label{eq:xi1_p}
\end{align}
and
\begin{align}
\ket{\xi_{2},p} =&  
\tfrac{1}{\sqrt{N_{\rm uc}+1/2}}
\left[
\sin \left( \tfrac{(n_{\rm uc}-\frac{1}{2}) \pi p }{N_{\rm uc} + \frac{1}{2}} \right)
\sqrt{ 1 - \tfrac{J}{\ep_p} } \ket{ \chi_\alp^- } \right.
\nonumber \\
&+
\left.
\sin \left( \tfrac{n_{\rm uc} \pi p }{N_{\rm uc} + \frac{1}{2}} \right)
\sqrt{ 1 + \tfrac{J}{\ep_p} } \ket{ \chi_\bb^- } 
\right]
\label{eq:xi2_p}
\end{align}
where $\ep_p =  \sqrt{J^2 + 4t_0^2 \cos^2 \left( \tfrac{p \pi}{2N_{\rm uc} + 1} \right)}$.

For the non-equilibrium Green function analysis of the transmission, we
will work in the basis of the eigenstates 
of the lower band of the isolated $\xh$-AFM region. 
The presence of the Fano resonances with zeros at the energies $-\ep_p$ 
guides us to focus on the degenerate $2 \times 2$ $p$ subspace 
defined by the two degenerate states $\ket{\xi_{1},p}$ and $\ket{\xi_{2},p}$.
We will consider how the the coupling to the anti-aligned, $\pm \zh$-FM contacts
affects the degenerate $p$ subspace and determine the transmission resulting from
these two states.
In the FM contacts we use the localized orbital $\ket{\uparrow}$,
$\ket{\downarrow}$ basis. 
The basis states at site $0$, the last site of the left contact, are denoted as
$\{ \ket{0,\uparrow}, \ket{0,\downarrow} \}$
and at site $N+1$, the first site of the right contact, 
they are denoted as
$\{ \ket{N+1,\uparrow}, \ket{N+1,\downarrow} \}$.

The self energies due to coupling of the AFM device states to the up spin band of the left contact and the
down spin band of the right contact require the surface Green functions and the matrix elements.
The Hamiltonian matrix element between $\ket{\xi_1,p}$ and the 
spin up band of the left contact is
\begin{align}
t_{0 \uparrow,1} =& \bra{0, \uparrow} H \ket{\xi_1,p} 
\nonumber \\
=&
\tfrac{1}{\sqrt{N_{\rm uc}+1/2}} 
\sin \left( \tfrac{\frac{1}{2} \pi p }{N_{\rm uc} + \frac{1}{2}} \right)
\sqrt{ 1 + \tfrac{J}{\ep_p} }
 \bra{0, \uparrow} H \ket{ \chi_\alp^+ } .
\end{align}
The matrix element
$\bra{0, \uparrow} H \ket{ \chi_\alp^+ } = t_0 / \sqrt{2}$, 
so that 
\begin{equation}
t_{0 \uparrow,1} = \tfrac{1}{\sqrt{2N_{\rm uc}+1}} 
\sin \left( \tfrac{\frac{1}{2} \pi p }{N_{\rm uc} + \frac{1}{2}} \right)
\sqrt{ 1 + \tfrac{J}{\ep_p} } \: t_0 .
\label{eq:t1up}
\end{equation}
The matrix element between $\ket{\xi_2,p}$ and the spin up band of the left contact is 
\begin{align}
t_{0 \uparrow,2} &= \bra{0, \uparrow} H \ket{\xi_{2},p} 
\nonumber \\
&= 
\tfrac{1}{\sqrt{N_{\rm uc}+1/2}} 
\sin \left( \tfrac{\frac{1}{2} \pi p }{N_{\rm uc} + \frac{1}{2}} \right)
\sqrt{ 1 - \tfrac{J}{\ep_p} }
 \bra{0, \uparrow} H \ket{ \chi_\alp^- }
\nonumber \\
&=
\tfrac{1}{\sqrt{2N_{\rm uc}+1}} 
\sin \left( \tfrac{\frac{1}{2} \pi p }{N_{\rm uc} + \frac{1}{2}} \right)
\sqrt{ 1 - \tfrac{J}{\ep_p} } \: t_0 .
\label{eq:t2up}
\end{align}
The nonzero matrix elements of $\ket{\xi_1,p}$ and $\ket{\xi_2,p}$ to the 
spin down band of the right contact are
\begin{align}
t_{1, N+1 \downarrow} =& \bra{ \xi_{1},p } H \ket{N+1, \downarrow}
\nonumber \\
=&
\tfrac{1}{\sqrt{N_{\rm uc}+1/2}} 
\sin \left( \tfrac{N_{\rm uc} \pi p }{N_{\rm uc}+\frac{1}{2}} \right)
\sqrt{ 1 - \tfrac{J}{\ep_p} }
\bra{ \chi_\bb^+ } H \ket{N+1, \downarrow} 
\nonumber \\
=& 
\tfrac{1}{\sqrt{2N_{\rm uc}+1}} 
\sin \left( \tfrac{N_{\rm uc}  \pi p }{N_{\rm uc} + \frac{1}{2}} \right)
\sqrt{ 1 - \tfrac{J}{\ep_p} } \: t_0 ,
\label{eq:t1dn}
\end{align}
and
\begin{align}
t_{2, N+1 \downarrow} =& \bra{ \xi_{2},p } H \ket{N+1, \downarrow}
\nonumber \\
=&
\tfrac{1}{\sqrt{N_{\rm uc}+1/2}} 
\sin \left( \tfrac{N_{\rm uc} \pi p }{N_{\rm uc}+\frac{1}{2}} \right)
\sqrt{ 1 + \tfrac{J}{\ep_p} }
\bra{ \chi_\bb^- } H \ket{N+1, \downarrow} 
\nonumber \\
=& 
- \tfrac{1}{\sqrt{2N_{\rm uc}+1}} 
\sin \left( \tfrac{N_{\rm uc}  \pi p }{N_{\rm uc} + \frac{1}{2}} \right)
\sqrt{ 1 + \tfrac{J}{\ep_p} } \: t_0 .
\label{eq:t2dn}
\end{align}
To shorten the notation, we define $t_1 \equiv  t_{0\uparrow,1}$, and the ratio
\begin{equation}
\mu \equiv \sqrt{\tfrac{1-J/\ep_p}{1+J/\ep_p}}.
\label{eq:mu}
\end{equation}
Furthermore, since 
$\sin \! \left( \tfrac{N_{\rm uc} \pi p }{N_{\rm uc}+\frac{1}{2}} \right) = 
(-1)^{p+1} \sin\!  \left( \tfrac{\frac{1}{2} \pi p }{N_{\rm uc} + \frac{1}{2}} \right)$, 
the 4 matrix elements are related as follows,
\begin{align}
t_{0\uparrow,1}&\equiv t_1
\nonumber \\
t_{0\uparrow,2}&= \mu t_1
\nonumber \\
t_{1,N+1\downarrow}&= (-1)^{p+1} \mu t_1
\nonumber \\
t_{2,N+1\downarrow}&= (-1)^{p} t_1 .
\label{eq:t_relations}
\end{align}

In the FM leads, the surface Green function is diagonal in the spin.
Within the $2\times 2$ degenerate subspace, the self energies due to coupling to
the lower spin-polarized bands of the left and right contacts are
\begin{equation}
\Sigma_{i,j}^{\mathcal L} = t_{i,0\uparrow} g_{0\uparrow , 0\uparrow} t_{0\uparrow,j}
\label{eq:sigL_general}
\end{equation}
and
\begin{equation}
\Sigma_{i,j}^{\mathcal R} = t_{i,N+1 \downarrow}g_{N+1\downarrow , N+1\downarrow} t_{N+1\downarrow,j},
\label{eq:sigR_general}
\end{equation}
respectively.
Since the bands of the left and right contacts are anti-aligned, the surface
Green functions are equal, 
\begin{equation}
g^s \equiv g_{0\uparrow , 0\uparrow} = g_{N+1\downarrow , N+1\downarrow}, 
\label{eq:gs_def}
\end{equation}
and given by
\begin{equation}
g^s= 
\frac{1}{|t_0|} \left[ \frac{E + J}{2|t_0|} - i \sqrt{1 - \left( \tfrac{E+J}{2t_0} \right)^2 } \right]
\label{eq:gs}
\end{equation}
where $(-J - 2|t_0| \le E \le -J + 2|t_0| )$.
Although it will not be needed, for energies 
below the bottom of the band $(E < -J - 2|t_0|)$, 
and above the top of the band 
$(-J+2|t_0| < E)$,
the surface Green function is purely real and given by
$g^s = \frac{E+J}{2t^2_0}  \left[ 1  - \sqrt{1 - \left( \frac{2t_0}{E+J} \right)^2} \right]$.

We are now ready to construct the self energies and Green function of the device within the 
$2\times 2$ degenerate subspace.
The self energy matrix due to coupling to the left lead is
\begin{equation}
\Sigma^{\mathcal L} =
\left[
\begin{array}{cc}
1 & \mu\\ 
\mu  & \mu^2 \\ 
\end{array}
\right] t_1^2 g^s  ,
\label{eq:SigLmatrix}
\end{equation}
and
\begin{equation}
\Gamma^{\mathcal L} =
\left[
\begin{array}{cc}
1 & \mu\\ 
\mu  & \mu^2 \\ 
\end{array}
\right] t_1^2 a^s  .
\label{eq:GLmatrix}
\end{equation}
Similarly, the self energy matrix due to coupling to the right lead is
\begin{equation}
\Sigma^{\mathcal R} =
\left[
\begin{array}{cc}
\mu^2 & - \mu \\ 
- \mu & 1 \\ 
\end{array}
\right] t_1^2 g^s  ,
\label{eq:SigRmatrix}
\end{equation}
and
\begin{equation}
\Gamma^{\mathcal R} =
\left[
\begin{array}{cc}
\mu^2 & - \mu \\ 
- \mu & 1 \\ 
\end{array}
\right] t_1^2 a^s  .
\label{eq:GRmatrix}
\end{equation}
In Eqs. (\ref{eq:SigLmatrix}) - (\ref{eq:GRmatrix}), 
we utilized Eqs. (\ref{eq:t_relations}) - (\ref{eq:gs_def}) and the relation $a^s = -2 \: {\rm Im}\{ g^s \}$
where $a^s$ is the surface spectral function.

The Green function of the $2\times 2$ degenerate $p$ subspace is
\begin{align}
G &= \left[ E - H_D - \Sigma^{\mathcal L} - \Sigma^{\mathcal R} \right]^{-1}
\nonumber \\
&=
\left[
\begin{array}{cc}
E + \ep_p - (1 + \mu^2) t_1^2 g^s & 0\\
0 & E + \ep_p - ( 1 + \mu^2 ) t_1^2 g^s
\end{array}
\right]^{-1}
\nonumber \\
&=
g
\left[
\begin{array}{cc}
1 & 0\\
0 & 1 
\end{array}
\right]
\end{align}
where $g = [E + \ep_p - ( 1 + \mu^2 ) t_1^2 g^s]^{-1}$.
The transmission is 
$T = {\rm tr} \{ \Gamma^{\mathcal L} G \Gamma^{\mathcal R} G^\dagger \}$,
and since $G$ is diagonal and proportional to the identity matrix, this becomes
\begin{equation}
T = |g|^2 {\rm tr} \{ \Gamma^{\mathcal L} \Gamma^{\mathcal R} \} .
\label{eq:T_tr2}
\end{equation}
However, $\Gamma^{\mathcal L}$ and $\Gamma^{\mathcal R}$ are degenerate, and the matrix product
$\Gamma^{\mathcal L} \Gamma^{\mathcal R} = 0$.
Thus, the transmission is identically equal to $0$ when the $\xh$-AFM region is directly
coupled to the two anti-aligned $\pm \zh$-FM contacts.

This result does not depend on the exact cancellation of the off-diagonal elements of the 
self energies.
If we detune $J$ between the left and right contacts so that the left
$g^s$ differs from the right $g^s$,
or if we introduce asymmetry in the coupling by replacing $t_0$ with a different
hopping element $t_{\mathcal L}$ in 
Eqs. (\ref{eq:t1up}) and (\ref{eq:t2up}),
the transmission does not increase.
Thus, the zero transmission does not depend on a perfect cancellation of the off diagonal
elements of the sum $\Sigma^{\mathcal L} + \Sigma^{\mathcal R}$.

To obtain a transmission with a Fano resonance, the ratio of the hopping
elements to the left contact $t_{0 \uparrow,2}/t_{0 \uparrow,1} = \mu$ 
must be different 
from the ratio of the hopping elements to the right contact,
$-t_{1, N+1 \downarrow} / t_{2, N+1 \downarrow}$. 
To show this analytically, 
we allow for modification of the ratio $\mu$,
by defining new ratios $u$ and $v$ via
\begin{equation}
t_{0 \uparrow,2} \equiv u t_{0 \uparrow,1} = u t_1
\label{eq:u_def}
\end{equation}
and
\begin{equation}
-t_{1,N+1 \downarrow} \equiv  v t_{2,N+1 \downarrow} = (-1)^p \: v t_1 .
\label{eq:v_def}
\end{equation}
To mimick reduced coupling due to tunneling,
we replace $t_1$ on the right hand side of Eq. (\ref{eq:u_def}) 
with $t^{\mathcal L} = \zeta t_1$ with $\zeta < 1$.
We set the surface Green function to be a constant and only consider
the imaginary part, $g^s = -i a^s/2$.
The self energies of Eqs. (\ref{eq:SigLmatrix}) - (\ref{eq:GRmatrix}) become
\begin{equation}
\Sigma^{\mathcal L} =
\left[
\begin{array}{cc}
1 &  u  \\ 
u  &  u^2
\end{array}
\right] \zeta^2 t_1^2 (-i a^s/2) 
\label{eq:SigLmatrix2}
\end{equation}
\begin{equation}
\Gamma^{\mathcal L} =
\left[
\begin{array}{cc}
1 &  u  \\ 
u  &  u^2
\end{array}
\right] \zeta^2 t_1^2 a^s  
\label{eq:GLmatrix2}
\end{equation}
\begin{equation}
\Sigma^{\mathcal R} =
\left[
\begin{array}{cc}
v^2 & -v \\
-v & 1
\end{array}
\right] t_1^2 (-i a^s/2) 
\label{eq:SigRmatrix2}
\end{equation}
\begin{equation}
\Gamma^{\mathcal R} =
\left[
\begin{array}{cc}
v^2 & -v \\
-v & 1
\end{array}
\right] t_1^2 a^s  .
\label{eq:GRmatrix2}
\end{equation}
The Green function then becomes (with $\zeta=1$)
\begin{equation}
G = 
\left[
\begin{array}{cc}
E + \ep_p  + i \tfrac{1}{2} a t_1^2(1+v^2) & i\tfrac{1}{2} a t_1^2(u-v) \\
i\tfrac{1}{2} a t_1^2(u-v) & E + \ep_p  + i \tfrac{1}{2} a t_1^2(1+u^2) 
\end{array}
\right]^{-1}
\label{eq:G_w.as.only}
\end{equation}
and the transmission is
\begin{align}
T& = {\rm tr} \{ \Gamma^{\mathcal L} G \Gamma^{\mathcal R} G^\dagger \} 
\nonumber \\
&=
\tfrac{16 t_1^4 a^2 (u-v)^2 (E+\ep_p)^2}{
\left[ 4(E+\ep_p)^2 - t_1^4 a^2 (1+uv)^2\right]^2 +
4^2t_1^4 (E+\ep_p)^2 (2+u^2+v^2)^2} \: .
\label{eq:Fano_analytical}
\end{align}
Two important points to take away from this are that
the zero in the numerator occurs exactly at the bound state energy,
and the transmission is identically equal to zero if the coupling ratios
$u$ and $v$ are equal.

Two poles of the transmission are
\begin{equation}
E = -\ep_p - i 
\tfrac{a t_1^2}{4} 
\left[ 2\!+\!u^2\!+\!v^2 \pm 
|u\!-\!v| \sqrt{4\!+\!(u\!+\!v)^2}
\right]
\end{equation}
and the other two poles are their complex conjugates.
The real parts of all of the poles are at the same energy as the zero.
Thus, this transmission contains the zero, but 
it does not have the analytic form of a Fano resonance. 
For a Fano resonance, the pole must be shifted slightly away from the zero.
This occurs when the degeneracy of the two states $\ket{\xi_1,p}$ and $\ket{\xi_2,p}$
is broken.

Breaking of the degeneracy occurs when the AFM layer couples to a normal layer,
either metallic or insulating.
This seems counterintuitive, since the two states are 100\% spin polarized and anti-aligned,
so one would expect that 
a finite magnetic moment or magnetic field would be required
to break the degeneracy.
Two things work to break the degeneracy.
The $\alpha$ site of the left most AFM layer couples most strongly to the normal layer on the left.
In our nearest neighbor tight binding model, it is the only site that couples
to the normal layer.
On the $\alpha$ site, the $\xh$ spin is weighted more heavily than the
$-\xh$ spin, and
the ratio of the two different weights is $\mu$.
This results in different couplings of the two degenerate
states $\ket{\xi_1,p}$ and $\ket{\xi_2,p}$ to the normal metal.

To see this, consider a normal state $\ket{\psi_0}$
with energy $\ep_0 = 0$ coupled on the left to the two degenerate states
$\ket{\xi_1,p}$ and $\ket{\xi_2,p}$ and work entirely within the
$\pm \xh$ spin basis so that the spin basis of state $\psi_0$
is $\left\{ \bm{\chi}_0^+, \bm{\chi}_0^- \right\} = \left\{ \frac{1}{\sqrt{2}}[1 \: 1]^T, \; \frac{1}{\sqrt{2}}[1 \: \text{-}1]^T \right\}$.
The nonzero matrix elements that couple the two degenerate AFM states to the normal 
state are
\begin{align}
t_{\chi_0^+,1}&= \bra{\chi_0^+}H\ket{\xi_1,p} 
\nonumber \\
&=  \tfrac{1}{\sqrt{N_{\rm uc}+1/2}} 
\sin \left( \tfrac{\frac{1}{2} \pi p }{N_{\rm uc} + \frac{1}{2}} \right)
\sqrt{ 1 + \tfrac{J}{\ep_p} }
 \bra{\chi_0^+ } H \ket{ \chi_\alp^+ }
\nonumber \\
&=
\tfrac{1}{\sqrt{N_{\rm uc}+1/2}} 
\sin \left( \tfrac{\frac{1}{2} \pi p }{N_{\rm uc} + \frac{1}{2}} \right)
\sqrt{ 1 + \tfrac{J}{\ep_p} } t_0
\label{eq:t0+2}
\end{align}
and
\begin{align}
t_{\chi_0^-,2}&= \bra{\chi_0^-}H\ket{\xi_2,p} 
\nonumber \\
&=  \tfrac{1}{\sqrt{N_{\rm uc}+1/2}} 
\sin \left( \tfrac{\frac{1}{2} \pi p }{N_{\rm uc} + \frac{1}{2}} \right)
\sqrt{ 1 - \tfrac{J}{\ep_p} }
\bra{\chi_0^- } H \ket{ \chi_\alp^- }
\nonumber \\
&=
\tfrac{1}{\sqrt{N_{\rm uc}+1/2}} 
\sin \left( \tfrac{\frac{1}{2} \pi p }{N_{\rm uc} + \frac{1}{2}} \right)
\sqrt{ 1 - \tfrac{J}{\ep_p} } t_0
\label{eq:t0-2}
\end{align}
Defining $t_{0,1} \equiv t_{\chi_0^+,1}$, then $t_{\chi_0^-,2} = \mu t_{0,1}$, and
the  Hamiltonian matrix is
\begin{equation}
H = 
\left[
\begin{array}{cccc}
0 & 0 &       t_{0,1} & 0 \\
0 & 0 &       0 & \mu t_{0,1} \\
t_{0,1} & 0 & -\ep_p & 0 \\
0 & \mu t_{0,1}  &   0 & -\ep_p
\end{array}
\right] .
\end{equation}
The state $\ket{\xi_1,p}$ splits into two states with energies
$E_1 = -\ep_p/2 \pm \sqrt{\left(\frac{\ep_p}{2}\right)^2 + t_{0,1}^2}$
and the state $\ket{\xi_2,p}$ splits into two states with energies
$E_2 = -\ep_p/2 \pm \sqrt{\left(\frac{\ep_p}{2}\right)^2 + \mu^2 t_{0,1}^2}$.
The eigenstates evolving from $\ket{\xi_1,p}$ are
$\left[ \ket{\chi_0^+} \pm \ket{\xi_1,p}\right]/\sqrt{2}$,
and those evolving from $\ket{\xi_2,p}$ are
$\left[ \ket{\chi_0^-} \pm \ket{\xi_2,p}\right]/\sqrt{2}$.
The coupling of the degenerate AFM states to the normal state
lifts the spin degeneracy of the AFM states, 
and, 
since $\ket{\xi_1,p}$ is composed entirely of $\ket{\chi^+}$ spins
and $\ket{\xi_2,p}$ is composed entirely of $\ket{\chi^-}$ spins,
each state is spin polarized.

We now take into account splitting of the levels in the $2\times 2$
diagonal device Hamiltonian $H_D$ letting
$\ep_{1} = -\ep_p + \Delta$ and $\ep_{2} = -\ep_p - \Delta$
so that the Green function of Eq. (\ref{eq:G_w.as.only}) is modified to
\begin{equation}
\arraycolsep=-3pt\def\arraystretch{1.3}
G \!=\! 
\left[
\begin{array}{cc}
E\! +\! \ep_p\! -\! \Delta  + i \tfrac{1}{2} a t_1^2(1\!+\!v^2) & i\tfrac{1}{2} a t_1^2(u\!-\!v) \\
i\tfrac{1}{2} a t_1^2(u\!-\!v) & E\! +\! \ep_p\! + \!\Delta  + i \tfrac{1}{2} a t_1^2(1\!+\!u^2) 
\end{array}
\right]^{-1}
\end{equation}
and the resulting transmission is
\begin{widetext}
\begin{equation}
T = \frac{16 t_1^4 a^2 \left((E+\ep_p)(u-v) - (u+v)\Delta \right)^2}{
\left[ 4[ (E+\ep_p)^2 -\Delta^2] - t_1^4 a^2 (1+uv)^2 \right]^2 +
4a^2 t_1^4 \left[ (E+\ep_p) (2+u^2+v^2) - (u^2-v^2)\Delta \right]^2} \: .
\end{equation}
\end{widetext}
For $u \ne v$, the zero is shifted to $-\ep_p + \Delta \frac{u+v}{u-v}$. 
When $u=v$, the zero disappears, but the transmission is still finite
and proportional to $\Delta^2$.
The poles are not easy to interpret analytically, therefore, we numerically evaluate
the transmisison.

\begin{figure}[t]
\includegraphics[width=3.2in]{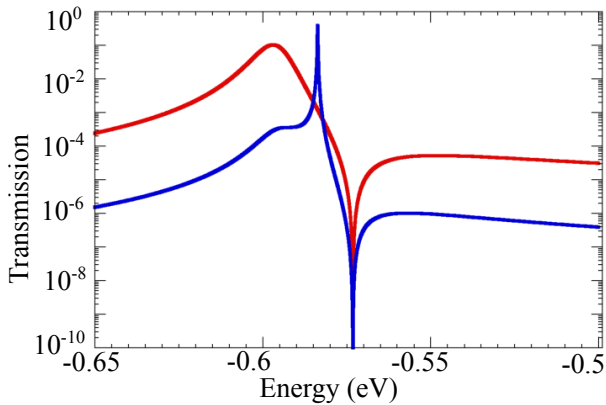}
\caption{
Transmission resulting from two spin degenerate $\xh$-AFM states coupled to
anti-aligned $\pm \zh$-FM contacts.
The parameters used are $N_{uc} = 8$, $t_0 = 0.38$ eV, $J = 0.15$ eV, $\Delta = 0.0005$ eV,
$u = 1.1 \mu$, $v = \mu$, and $p=4$.
For the upper red curve, $\zeta = 1$, and for the lower blue curve, $\zeta = 0.1$.
}
\label{fig:Fanos_analytical}
\end{figure}

To demonstrate that the above analysis of the transmission is valid,
we evaluate the transmission
using the self energies from Eqs. (\ref{eq:SigLmatrix2}) - (\ref{eq:GRmatrix2}) 
with the approximate
surface Green functions and spectral functions replaced with the exact
expressions from Eq. (\ref{eq:gs}). 
We allow for splitting of the degenerate energies, $-\ep_p \rightarrow -\ep_p \pm \Delta$,
and scaling of $t_1$ ($\zeta < 1$) in Eqs. (\ref{eq:SigLmatrix2})-(\ref{eq:GLmatrix2}) 
to mimick a tunnel barrier.
Fig. \ref{fig:Fanos_analytical} shows the transmission calculated from the degenerate subspace
corresponding to the $p=4$ state of an 8 unit cell $\xh$-AFM coupled to anti-aligned
$\pm \zh$-FM contacts. 
When $u = v = \mu$, the transmission is below $10^{-15}$. 
For both curves shown, the coupling ratio of the hopping elements to
the left contact are changed to $u=1.1 \mu$, and 
a $\Delta = 0.5$ meV splitting of the energies is included.
The upper curve results when $\zeta = 1$ in the self-energy expressions
(\ref{eq:SigLmatrix2}) - (\ref{eq:GRmatrix2}).
The lower blue curve results when $\zeta = 0.1$ in Eqs. (\ref{eq:SigLmatrix2}) and (\ref{eq:GLmatrix2})
to approximate reduced coupling to the left contact.

Thus, the essential physics giving rise to the Fano resonances is well explained 
by the symmetry of the couplings of the anti-aligned $\pm \zh$-FM contacts 
to the spin degenerate subspace formed by the 
$\pm \xh$ polarized states $\ket{\xi_1,p}$ and $\ket{\xi_2,p}$ and by the
breaking of that symmetry and degeneracy 
due to the presence of a normal metal or tunnel barrier layer.
In the absence of such a layer, the symmetry and degeneracy is preserved,
and the $T_{\uparrow , \downarrow}$ transmission channel is blocked.
The insertion of a normal layer between the top FM contact and the AFM serves both to modify
the ratio $\mu$ of the matrix elements, since $J=0$ in the normal region, and to break the spin
degeneracy of the AFM states.

\section{Magneto tunneling detection of ferromagnetic skyrmions}
\label{sec:3D_TMR}

In this section, we numerically study the effect of a skyrmion on the tunneling magnetoresistance (TMR) of the MTJ structure shown in Fig. \ref{fig:AFMSchematic}. 
The goal is to calculate the tunneling magnetoresistance in the presence of and in the absence of a skyrmion. 
The device region, in the absence of a skyrmion, consists of either a FM or 
AFM with with magnetization or N\'{e}el vector oriented in the $\hat{z}$ direction. 
The results in this section are identical for both Bloch-type and N\'{e}el-type skyrmions. 

First, we insert a FM skyrmion (FM-SK) into the device region, the green layer in Fig. \ref{fig:AFMSchematic}, 
and compare the TMR with a $\hat{z}$ FM device.
The length of the device region is 2 nm, which consists of 4 atomic layers, and
the barrier thickness is 1 nm. 
The exchange potential of the top contact is set to be $J = 500$ meV so that the injected current is 100\% spin-polarized \cite{Bowen2005}.  
The exchange potential of the device region is set to be $J_{device} = 120$ meV, to make sure that 
the there is only one ferromagnetic band at the Fermi level ($E_f = -2.1$ eV). 
The value of $\eta$ used in the iterative calculation the surface Green functions is 0.1 meV in all calculations.  

\begin{figure}[H]
\centering 
\begin{subfigure}[b]{0.5\textwidth}
         \centering
         \includegraphics[width=\textwidth]{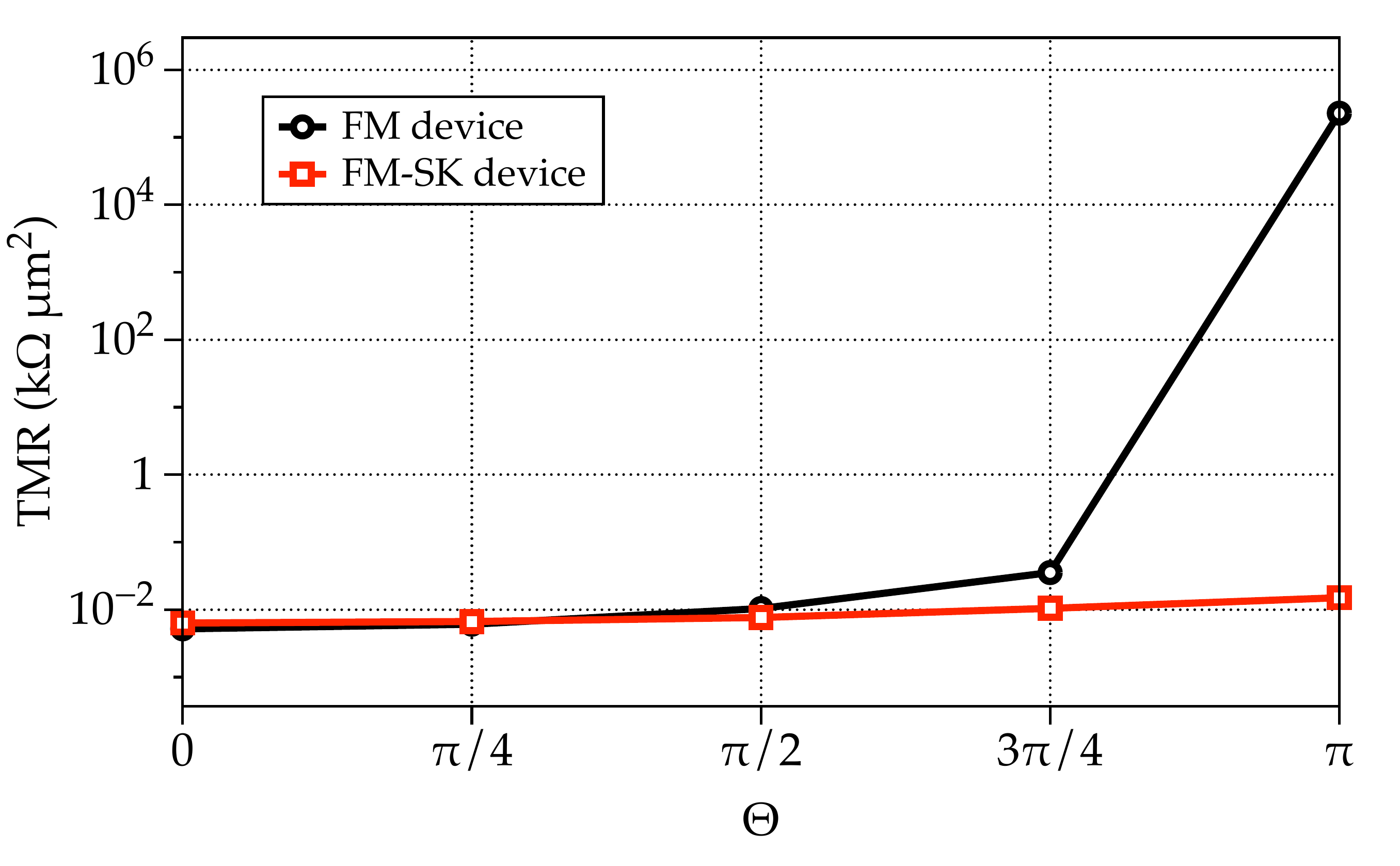} 
         \caption{}
         \label{fig:TMR_FM}
     \end{subfigure}
     \hfill
\begin{subfigure}[b]{0.5\textwidth}
         \centering
         \includegraphics[width=\textwidth]{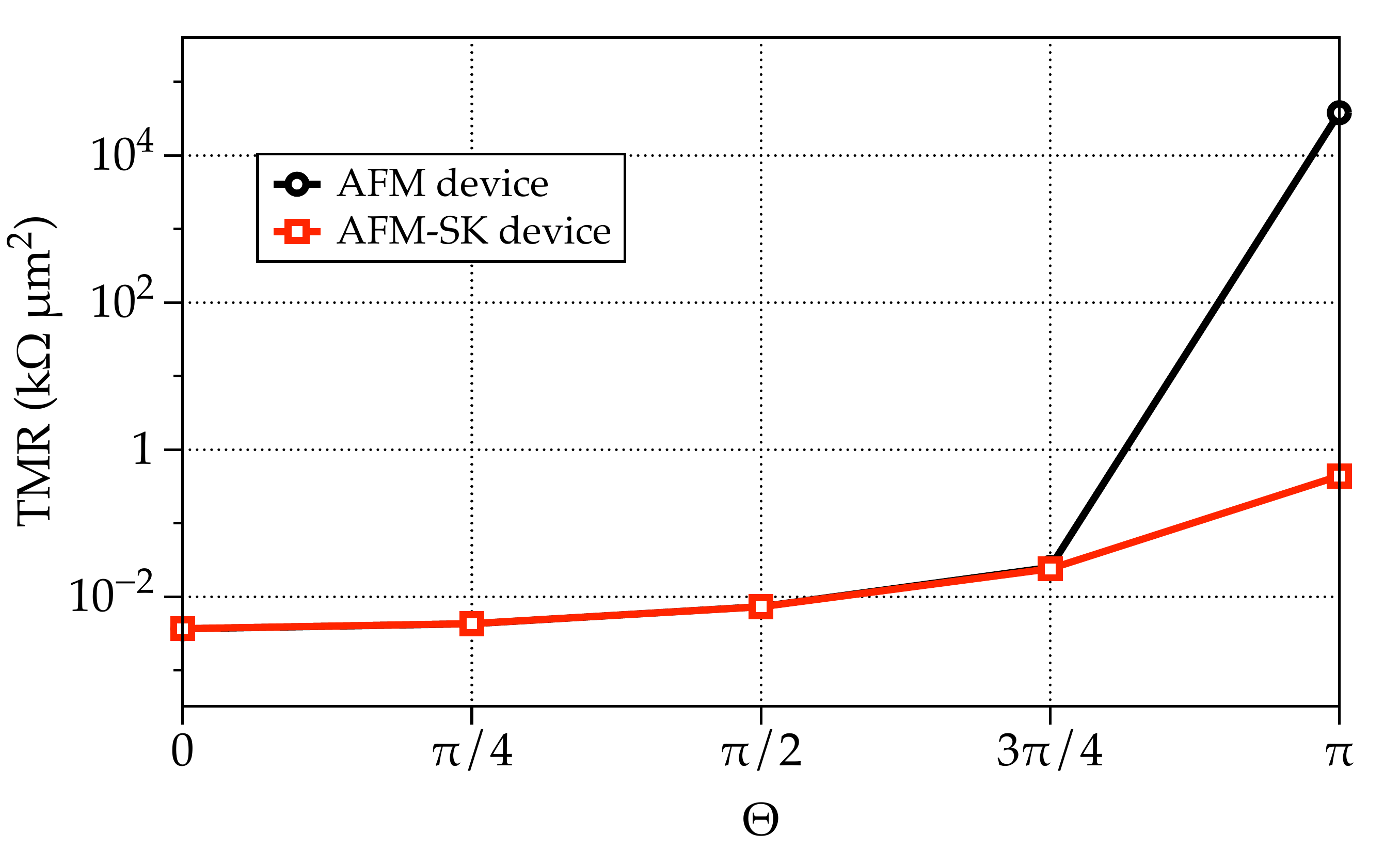} 
         \caption{}
         \label{fig:TMR_AFM}
     \end{subfigure}
 \hfill
\begin{subfigure}[b]{0.5\textwidth}
         \centering
         \includegraphics[width=\textwidth]{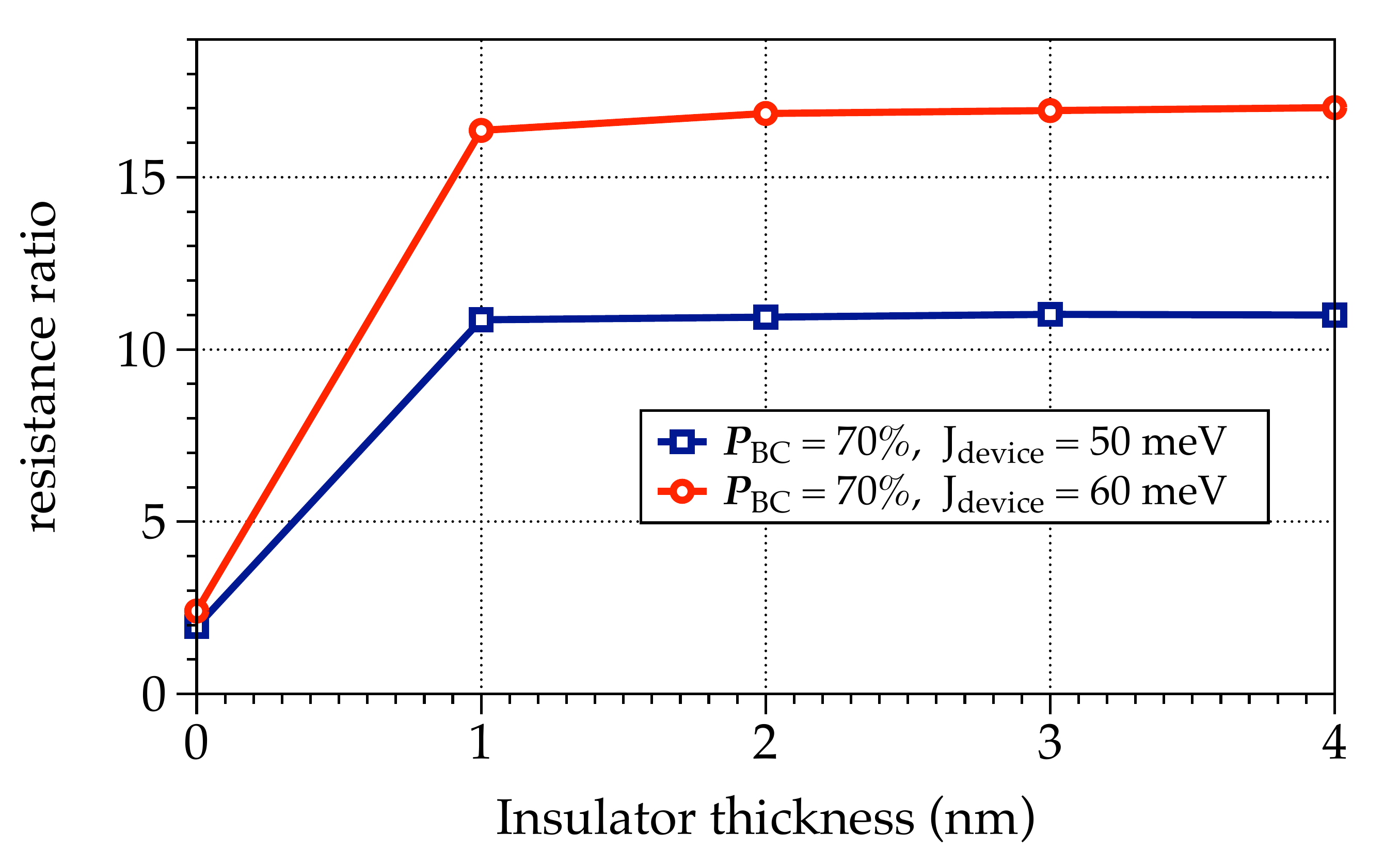} 
         \caption{}
         \label{fig:ratio_W}
\end{subfigure}
\caption{%
TMR vs. $\Theta$ for (a) a ferromagnetic (FM) device and FM skyrmion (FM-SK) device, 
and (b) an antiferromagnetic (AFM) device and AFM skyrmion (AFM-SK) device.
The exchange potentials of the device layer and contacts are 120 meV and 500 meV, respectively. 
The insulating barrier thickness is 1 nm.
(c) Resistance ratio versus insulator thickness at $\Theta=\pi$ for two different values of 
the device exchange potential as shown in the legend. The bottom contact is $70\%$ polarized.
}
\label{fig:AFM_SKX_R}
\end{figure}

The TMR in the presence (FM-SK) and absence (FM) of a FM skyrmion is shown in Fig. \ref{fig:TMR_FM} 
as a function of the angle $\Theta$ between the magnetic moments of the two FM contacts. 
When $\Theta = 0$, the magnetic moments of the contacts are aligned, and when $\Theta = \pi$, they are anti-aligned. 
As $\Theta$ increases from 0 to $\pi$, the resistance of the $\zh$ FM device increases from 8 $\Omega \: \mu$m$^2$ to
2 $\times$ $10^5$  k$\Omega \: \mu$m$^2$.
However, the TMR change in the presence of a FM-SK is negligible. 
For the $\zh$ polarized FM in the device region, the TMR depends strongly on $\Theta$ as one would expect for a typical MTJ, 
since the magnetic moment of the FM in the device layer is aligned with the magnetic moment of the bottom contact. 
For the $\zh$ FM device at $\Theta = \pi$, the spin-flip conduction channel, 
which is the primary conduction channel available with good polarization of the contacts, is blocked.

For the FM skyrmion device, the majority of the spins in the central FM layer have an in-plane component, 
since the spin texture of the skyrmion covers the Bloch sphere.
The in-plane spin components of the FM skyrmion open the spin-flip conduction channel, 
and the resistance becomes insensitive to the polarization direction of the contacts.
The slight increase in the TMR of the FM-SK is due to the fact that the skyrmion does not cover the entire 
pillar area, and also its spin texture has a mix of in-plane and out of plane components.
The TMR of an AFM device as a function of $\Theta$ is shown in Fig. \ref{fig:TMR_AFM} 
with (AFM-SK) and without (AFM) a skyrmion.  
For the uniform AFM with a N\'{e}el vector along the $\hat{z}$ axis in the device region,
the TMR depends strongly on $\Theta$, 
and it is similar to the $\hat{z}$ polarized FM device of Fig. \ref{fig:TMR_FM}.
At $\Theta = \pi$, the TMR of the $\hat{z}$ AFM device increases from 8 $\Omega \: \mu$m$^2$ to $8 \times 10^4$ k$\Omega \: \mu$m$^2$, 
and that of the AFM-SK device 
increases from 8 $\Omega \: \mu$m$^2$ at $\Theta = 0$ to 900 $\Omega \: \mu$m$^2$ at $\Theta = \pi$. 
With the anti-aligned FM contacts, the presence or absence of an AFM skyrmion changes the TMR by a factor of $9 \times 10^4$.

This resistance change in the presence or absence of an AFM skyrmion allows electrical detection, 
and the utility of such a detection scheme depends on the ratio of the resistance change.
We define the resistance ratio as the ratio of the TMR in the absence of the skyrmion 
to the TMR in the presence of the skyrmion ($R_{\rm AFM} / R_{\rm AFM-SK}$). 
Fig. \ref{fig:ratio_W}, shows calculations of the resistance ratio of an AFM device with anti-aligned FM contacts 
as a function of the insulator barrier thickness $W_I$ for 
finite contact polarization and reduced Hund's rule exchange coupling.
The polarization of the top contact remains at 100\% and the polarization of the bottom contact is reduced to $70\%$. 
Results are shown for two values of $J_{device} = 50$ and 60 meV.
AT $W_I=0$, (no tunnel barrier) the resistance ratio of both $J_{device} = 50$ and 60 meV is approximately 2. 
Since the bottom contact is 70\% polarized, there is a $\sim 30\%$ $T_{\uparrow, \uparrow}$ channel available
for current to flow in the $\zh$ AFM device, even though the FM contacts are anti-aligned.
For the AFM skyrmion device, the suppression of the $T_{\uparrow, \downarrow}$
transmission channel in the absence of a barrier does not occur,
since the AFM layers coupling to the contacts
contain spins of all angles, and the relationships given in Eq. (\ref{eq:t_relations}) do not hold.
The presence of an insulating barrier increases the resistance ratio.
For finite barrier thicknesses of 1 nm or more, the resistance ratio becomes independent of the barrier thickness, 
and for values of $J_{device} = 50$ and 60 meV, the ratios are 11 and 17, respectively. 

\section{Summary \& Conclusions}
\label{sec:conclusions}
Transport through a layered antiferromagnet between two FM contacts has been analyzed.
An ideal layered $\xh$-polarized AFM between two antialigned $\pm \zh$ FM contacts  
carries no current due to the $\pi$ phase difference of the matrix elements
coupling the spin degenerate states to the left and right contacts.
The ratio of the matrix elements of the $\pm \xh$ AFM spin states to one $\zh$ FM polarized contact 
is $\mu = \sqrt{\frac{1-J\ep_p}{1+J/\ep_p}}$,
and the ratio to the other oppositely polarized FM contact is $-\mu$. 
Inserting a normal metal layer or tunnel barrier layer between the top contact
and the AFM alters this ratio, since $J=0$ in a normal region, 
and it also breaks the degeneracy between the two spin states. 
The altering of the ratio $\mu$ allows transmission with a zero
at the bound state energy, and the breaking of the degeneracy shifts
the energy of the pole from the zero giving rise to a Fano resonance. 
This MTJ geometry with two antialigned $\pm \zh$ FM 
contacts can be used to sense an AFM skyrmion in a central AFM layer.
When the N\'{e}el vector of the AFM layer is vertically aligned, resistance is high.
The presence of an AFM skyrmion introduces in-plane components to the spin 
that act as an analogue of the oblique polarizer in the triple polizer experiment.
The $T_{\uparrow, \downarrow}$ transmission channel opens, and the resistance decreases. 
The resistance ratio ranges from 11 to 17 in the presence of a insulating barrier as 
the device exchange potential is changed from 50 to 60 meV for a $70\%$ polarized bottom contact.
\\

\noindent
{\bf Acknowledgements}: This work was supported as part of Spins and Heat in Nanoscale Electronic Systems (SHINES) an 
Energy Frontier Research Center funded by the U.S. Department of Energy, Office of Science, 
Basic Energy Sciences under Award \#DE-SC0012670. 
 
\newpage

\bibliography{APSBIB,BIBLIOGRAPHY_v2}

\begin{thebibliography}{42}%
\makeatletter
\providecommand \@ifxundefined [1]{%
 \@ifx{#1\undefined}
}%
\providecommand \@ifnum [1]{%
 \ifnum #1\expandafter \@firstoftwo
 \else \expandafter \@secondoftwo
 \fi
}%
\providecommand \@ifx [1]{%
 \ifx #1\expandafter \@firstoftwo
 \else \expandafter \@secondoftwo
 \fi
}%
\providecommand \natexlab [1]{#1}%
\providecommand \enquote  [1]{``#1''}%
\providecommand \bibnamefont  [1]{#1}%
\providecommand \bibfnamefont [1]{#1}%
\providecommand \citenamefont [1]{#1}%
\providecommand \href@noop [0]{\@secondoftwo}%
\providecommand \href [0]{\begingroup \@sanitize@url \@href}%
\providecommand \@href[1]{\@@startlink{#1}\@@href}%
\providecommand \@@href[1]{\endgroup#1\@@endlink}%
\providecommand \@sanitize@url [0]{\catcode `\\12\catcode `\$12\catcode
  `\&12\catcode `\#12\catcode `\^12\catcode `\_12\catcode `\%12\relax}%
\providecommand \@@startlink[1]{}%
\providecommand \@@endlink[0]{}%
\providecommand \url  [0]{\begingroup\@sanitize@url \@url }%
\providecommand \@url [1]{\endgroup\@href {#1}{\urlprefix }}%
\providecommand \urlprefix  [0]{URL }%
\providecommand \Eprint [0]{\href }%
\providecommand \doibase [0]{http://dx.doi.org/}%
\providecommand \selectlanguage [0]{\@gobble}%
\providecommand \bibinfo  [0]{\@secondoftwo}%
\providecommand \bibfield  [0]{\@secondoftwo}%
\providecommand \translation [1]{[#1]}%
\providecommand \BibitemOpen [0]{}%
\providecommand \bibitemStop [0]{}%
\providecommand \bibitemNoStop [0]{.\EOS\space}%
\providecommand \EOS [0]{\spacefactor3000\relax}%
\providecommand \BibitemShut  [1]{\csname bibitem#1\endcsname}%
\let\auto@bib@innerbib\@empty
\bibitem [{\citenamefont {Jungwirth}\ \emph {et~al.}(2016)\citenamefont
  {Jungwirth}, \citenamefont {Marti}, \citenamefont {Wadley},\ and\
  \citenamefont {Wunderlich}}]{AFM_spintronics_Jungwirth_NNano16}%
  \BibitemOpen
  \bibfield  {author} {\bibinfo {author} {\bibfnamefont {T.}~\bibnamefont
  {Jungwirth}}, \bibinfo {author} {\bibfnamefont {X.}~\bibnamefont {Marti}},
  \bibinfo {author} {\bibfnamefont {P.}~\bibnamefont {Wadley}}, \ and\ \bibinfo
  {author} {\bibfnamefont {J.}~\bibnamefont {Wunderlich}},\ }\bibfield  {title}
  {\enquote {\bibinfo {title} {Antiferromagnetic spintronics},}\ }\href
  {dx.doi.org/10.1038/nnano.2016.18} {\bibfield  {journal} {\bibinfo  {journal}
  {Nat. Nano}\ }\textbf {\bibinfo {volume} {11}},\ \bibinfo {pages} {231--241}
  (\bibinfo {year} {2016})}\BibitemShut {NoStop}%
\bibitem [{\citenamefont {O.}\ \emph {et~al.}(2017)\citenamefont {O.},
  \citenamefont {T.},\ and\ \citenamefont
  {J.}}]{2017_AFM_spintronics_Jungwirth_PSSR}%
  \BibitemOpen
  \bibfield  {author} {\bibinfo {author} {\bibfnamefont {Gomonay}\ \bibnamefont
  {O.}}, \bibinfo {author} {\bibfnamefont {Jungwirth}\ \bibnamefont {T.}}, \
  and\ \bibinfo {author} {\bibfnamefont {Sinova}\ \bibnamefont {J.}},\
  }\bibfield  {title} {\enquote {\bibinfo {title} {Concepts of
  antiferromagnetic spintronics},}\ }\href {\doibase 10.1002/pssr.201700022}
  {\bibfield  {journal} {\bibinfo  {journal} {physica status solidi (RRL) –
  Rapid Research Letters}\ }\textbf {\bibinfo {volume} {11}},\ \bibinfo {pages}
  {1700022} (\bibinfo {year} {2017})}\BibitemShut {NoStop}%
\bibitem [{\citenamefont {Baltz}\ \emph {et~al.}(2018)\citenamefont {Baltz},
  \citenamefont {Manchon}, \citenamefont {Tsoi}, \citenamefont {Moriyama},
  \citenamefont {Ono},\ and\ \citenamefont
  {Tserkovnyak}}]{2018_Tserkovnyak_RMP}%
  \BibitemOpen
  \bibfield  {author} {\bibinfo {author} {\bibfnamefont {V.}~\bibnamefont
  {Baltz}}, \bibinfo {author} {\bibfnamefont {A.}~\bibnamefont {Manchon}},
  \bibinfo {author} {\bibfnamefont {M.}~\bibnamefont {Tsoi}}, \bibinfo {author}
  {\bibfnamefont {T.}~\bibnamefont {Moriyama}}, \bibinfo {author}
  {\bibfnamefont {T.}~\bibnamefont {Ono}}, \ and\ \bibinfo {author}
  {\bibfnamefont {Y.}~\bibnamefont {Tserkovnyak}},\ }\bibfield  {title}
  {\enquote {\bibinfo {title} {Antiferromagnetic spintronics},}\ }\href
  {\doibase 10.1103/RevModPhys.90.015005} {\bibfield  {journal} {\bibinfo
  {journal} {Rev. Mod. Phys.}\ }\textbf {\bibinfo {volume} {90}},\ \bibinfo
  {pages} {015005} (\bibinfo {year} {2018})}\BibitemShut {NoStop}%
\bibitem [{\citenamefont {Grezes}\ \emph {et~al.}(2016)\citenamefont {Grezes},
  \citenamefont {Ebrahimi}, \citenamefont {Alzate}, \citenamefont {Cai},
  \citenamefont {Katine}, \citenamefont {Langer}, \citenamefont {Ocker},
  \citenamefont {Amiri},\ and\ \citenamefont
  {Wang}}]{KWang_ULowSwitchingAFM_APL16}%
  \BibitemOpen
  \bibfield  {author} {\bibinfo {author} {\bibfnamefont {C.}~\bibnamefont
  {Grezes}}, \bibinfo {author} {\bibfnamefont {F.}~\bibnamefont {Ebrahimi}},
  \bibinfo {author} {\bibfnamefont {J.~G.}\ \bibnamefont {Alzate}}, \bibinfo
  {author} {\bibfnamefont {X.}~\bibnamefont {Cai}}, \bibinfo {author}
  {\bibfnamefont {J.~A.}\ \bibnamefont {Katine}}, \bibinfo {author}
  {\bibfnamefont {J.}~\bibnamefont {Langer}}, \bibinfo {author} {\bibfnamefont
  {B.}~\bibnamefont {Ocker}}, \bibinfo {author} {\bibfnamefont {P.~Khalili}\
  \bibnamefont {Amiri}}, \ and\ \bibinfo {author} {\bibfnamefont {K.~L.}\
  \bibnamefont {Wang}},\ }\bibfield  {title} {\enquote {\bibinfo {title}
  {Ultra-low switching energy and scaling in electric-field-controlled
  nanoscale magnetic tunnel junctions with high resistance-area product},}\
  }\href {\doibase 10.1063/1.4939446} {\bibfield  {journal} {\bibinfo
  {journal} {Appl. Phys. Lett.}\ }\textbf {\bibinfo {volume} {108}},\ \bibinfo
  {pages} {012403} (\bibinfo {year} {2016})}\BibitemShut {NoStop}%
\bibitem [{\citenamefont {Dirac}(2010)}]{Dirac_PQM}%
  \BibitemOpen
  \bibfield  {author} {\bibinfo {author} {\bibfnamefont {P.~A.~M.}\
  \bibnamefont {Dirac}},\ }\href@noop {} {\emph {\bibinfo {title} {The
  Principles of Quantum Mechanics}}},\ \bibinfo {edition} {4th}\ ed.\ (\bibinfo
   {publisher} {Oxford University Press},\ \bibinfo {address} {Oxford},\
  \bibinfo {year} {2010})\BibitemShut {NoStop}%
\bibitem [{\citenamefont {N\'{u}nez}\ \emph {et~al.}(2006)\citenamefont
  {N\'{u}nez}, \citenamefont {Duine}, \citenamefont {Haney},\ and\
  \citenamefont {MacDonald}}]{MacDonald_AFM_Spin_Torques_PRB06}%
  \BibitemOpen
  \bibfield  {author} {\bibinfo {author} {\bibfnamefont {A.~S.}\ \bibnamefont
  {N\'{u}nez}}, \bibinfo {author} {\bibfnamefont {R.~A.}\ \bibnamefont
  {Duine}}, \bibinfo {author} {\bibfnamefont {Paul}\ \bibnamefont {Haney}}, \
  and\ \bibinfo {author} {\bibfnamefont {A.~H.}\ \bibnamefont {MacDonald}},\
  }\bibfield  {title} {\enquote {\bibinfo {title} {Theory of spin torques and
  giant magnetoresistance in antiferromagnetic metals},}\ }\href {\doibase
  10.1103/PhysRevB.73.214426} {\bibfield  {journal} {\bibinfo  {journal} {Phys.
  Rev. B}\ }\textbf {\bibinfo {volume} {73}},\ \bibinfo {pages} {214426}
  (\bibinfo {year} {2006})}\BibitemShut {NoStop}%
\bibitem [{\citenamefont {Haney}\ \emph {et~al.}(2007)\citenamefont {Haney},
  \citenamefont {Waldron}, \citenamefont {Duine}, \citenamefont {N\'{u}nez},
  \citenamefont {Guo},\ and\ \citenamefont
  {MacDonald}}]{Cr_Au_Cr_HGuo_MacDonald_PRB07}%
  \BibitemOpen
  \bibfield  {author} {\bibinfo {author} {\bibfnamefont {P.~M.}\ \bibnamefont
  {Haney}}, \bibinfo {author} {\bibfnamefont {D.}~\bibnamefont {Waldron}},
  \bibinfo {author} {\bibfnamefont {R.~A.}\ \bibnamefont {Duine}}, \bibinfo
  {author} {\bibfnamefont {A.~S.}\ \bibnamefont {N\'{u}nez}}, \bibinfo {author}
  {\bibfnamefont {H.}~\bibnamefont {Guo}}, \ and\ \bibinfo {author}
  {\bibfnamefont {A.~H.}\ \bibnamefont {MacDonald}},\ }\bibfield  {title}
  {\enquote {\bibinfo {title} {Ab initio giant magnetoresistance and
  current-induced torques in {Cr}/{Au}/{Cr} multilayers},}\ }\href {\doibase
  10.1103/PhysRevB.75.174428} {\bibfield  {journal} {\bibinfo  {journal} {Phys.
  Rev. B}\ }\textbf {\bibinfo {volume} {75}},\ \bibinfo {pages} {174428}
  (\bibinfo {year} {2007})}\BibitemShut {NoStop}%
\bibitem [{\citenamefont {Xu}\ \emph {et~al.}(2008)\citenamefont {Xu},
  \citenamefont {Wang},\ and\ \citenamefont
  {Xia}}]{STT_AFM_AR_FeMn_DFT_Xia_PRL08}%
  \BibitemOpen
  \bibfield  {author} {\bibinfo {author} {\bibfnamefont {Yuan}\ \bibnamefont
  {Xu}}, \bibinfo {author} {\bibfnamefont {Shuai}\ \bibnamefont {Wang}}, \ and\
  \bibinfo {author} {\bibfnamefont {Ke}~\bibnamefont {Xia}},\ }\bibfield
  {title} {\enquote {\bibinfo {title} {Spin-transfer torques in
  antiferromagnetic metals from first principles},}\ }\href {\doibase
  10.1103/PhysRevLett.100.226602} {\bibfield  {journal} {\bibinfo  {journal}
  {Phys. Rev. Lett.}\ }\textbf {\bibinfo {volume} {100}},\ \bibinfo {pages}
  {226602} (\bibinfo {year} {2008})}\BibitemShut {NoStop}%
\bibitem [{\citenamefont {Jia}\ \emph {et~al.}(2017)\citenamefont {Jia},
  \citenamefont {Tang}, \citenamefont {Wang},\ and\ \citenamefont
  {Qin}}]{Fe_MgO_FeMn_Cu_TAMR_PRB17}%
  \BibitemOpen
  \bibfield  {author} {\bibinfo {author} {\bibfnamefont {Xingtao}\ \bibnamefont
  {Jia}}, \bibinfo {author} {\bibfnamefont {Huimin}\ \bibnamefont {Tang}},
  \bibinfo {author} {\bibfnamefont {Shizhuo}\ \bibnamefont {Wang}}, \ and\
  \bibinfo {author} {\bibfnamefont {Minghui}\ \bibnamefont {Qin}},\ }\bibfield
  {title} {\enquote {\bibinfo {title} {Structure-dependent magnetoresistance
  and spin-transfer torque in antiferromagnetic
  $\mathrm{Fe}|\mathrm{MgO}|\mathrm{FeMn}|\mathrm{Cu}$ tunnel junctions},}\
  }\href {\doibase 10.1103/PhysRevB.95.064402} {\bibfield  {journal} {\bibinfo
  {journal} {Phys. Rev. B}\ }\textbf {\bibinfo {volume} {95}},\ \bibinfo
  {pages} {064402} (\bibinfo {year} {2017})}\BibitemShut {NoStop}%
\bibitem [{\citenamefont {Su}\ \emph {et~al.}(2019)\citenamefont {Su},
  \citenamefont {Zhang}, \citenamefont {L\"u}, \citenamefont {Hong},\ and\
  \citenamefont {You}}]{Mn3Pt_SrTiO3_Pt_TMR_PRAppl19}%
  \BibitemOpen
  \bibfield  {author} {\bibinfo {author} {\bibfnamefont {Yurong}\ \bibnamefont
  {Su}}, \bibinfo {author} {\bibfnamefont {Jia}\ \bibnamefont {Zhang}},
  \bibinfo {author} {\bibfnamefont {Jing-Tao}\ \bibnamefont {L\"u}}, \bibinfo
  {author} {\bibfnamefont {Jeongmin}\ \bibnamefont {Hong}}, \ and\ \bibinfo
  {author} {\bibfnamefont {Long}\ \bibnamefont {You}},\ }\bibfield  {title}
  {\enquote {\bibinfo {title} {Large magnetoresistance in an
  electric-field-controlled antiferromagnetic tunnel junction},}\ }\href
  {\doibase 10.1103/PhysRevApplied.12.044036} {\bibfield  {journal} {\bibinfo
  {journal} {Phys. Rev. Applied}\ }\textbf {\bibinfo {volume} {12}},\ \bibinfo
  {pages} {044036} (\bibinfo {year} {2019})}\BibitemShut {NoStop}%
\bibitem [{\citenamefont {Jia}\ \emph {et~al.}(2020)\citenamefont {Jia},
  \citenamefont {Cai},\ and\ \citenamefont {Jia}}]{Mn2Au_MTJ_ScChina20}%
  \BibitemOpen
  \bibfield  {author} {\bibinfo {author} {\bibfnamefont {Xing-Tao}\
  \bibnamefont {Jia}}, \bibinfo {author} {\bibfnamefont {Xiao-Lin}\
  \bibnamefont {Cai}}, \ and\ \bibinfo {author} {\bibfnamefont
  {Yu}~\bibnamefont {Jia}},\ }\bibfield  {title} {\enquote {\bibinfo {title}
  {Giant magnetoresistance in antiferromagnetic {Mn$_2$Au}-based tunnel
  junction},}\ }\href {\doibase 10.1007/s11433-019-1519-4} {\bibfield
  {journal} {\bibinfo  {journal} {SCIENCE CHINA Physics, Mechanics \&
  Astronomy}\ }\textbf {\bibinfo {volume} {accepted}},\ \bibinfo {pages} {--}
  (\bibinfo {year} {2020})}\BibitemShut {NoStop}%
\bibitem [{\citenamefont {Park}\ \emph {et~al.}(2011)\citenamefont {Park},
  \citenamefont {Wunderlich}, \citenamefont {Martí}, \citenamefont {Holý},
  \citenamefont {Kurosaki}, \citenamefont {Yamada}, \citenamefont {Yamamoto},
  \citenamefont {Nishide}, \citenamefont {Hayakawa}, \citenamefont {Takahashi},
  \citenamefont {Shick},\ and\ \citenamefont
  {Jungwirth}}]{TAMR_AFM_Jungwirth_2011}%
  \BibitemOpen
  \bibfield  {author} {\bibinfo {author} {\bibfnamefont {B.~G.}\ \bibnamefont
  {Park}}, \bibinfo {author} {\bibfnamefont {J.}~\bibnamefont {Wunderlich}},
  \bibinfo {author} {\bibfnamefont {X.}~\bibnamefont {Martí}}, \bibinfo
  {author} {\bibfnamefont {V.}~\bibnamefont {Holý}}, \bibinfo {author}
  {\bibfnamefont {Y.}~\bibnamefont {Kurosaki}}, \bibinfo {author}
  {\bibfnamefont {M.}~\bibnamefont {Yamada}}, \bibinfo {author} {\bibfnamefont
  {H.}~\bibnamefont {Yamamoto}}, \bibinfo {author} {\bibfnamefont
  {A.}~\bibnamefont {Nishide}}, \bibinfo {author} {\bibfnamefont
  {J.}~\bibnamefont {Hayakawa}}, \bibinfo {author} {\bibfnamefont
  {H.}~\bibnamefont {Takahashi}}, \bibinfo {author} {\bibfnamefont {A.~B.}\
  \bibnamefont {Shick}}, \ and\ \bibinfo {author} {\bibfnamefont
  {T.}~\bibnamefont {Jungwirth}},\ }\bibfield  {title} {\enquote {\bibinfo
  {title} {A spin-valve-like magnetoresistance of an antiferromagnet-based
  tunnel junction},}\ }\href {\doibase 10.1038/nmat2983} {\bibfield  {journal}
  {\bibinfo  {journal} {Nature Materials}\ }\textbf {\bibinfo {volume} {10}},\
  \bibinfo {pages} {347--351} (\bibinfo {year} {2011})}\BibitemShut {NoStop}%
\bibitem [{\citenamefont {Wang}\ \emph {et~al.}(2012)\citenamefont {Wang},
  \citenamefont {Song}, \citenamefont {Cui}, \citenamefont {Wang},
  \citenamefont {Zeng},\ and\ \citenamefont {Pan}}]{IrMn_TMR_Pan_PRL12}%
  \BibitemOpen
  \bibfield  {author} {\bibinfo {author} {\bibfnamefont {Y.~Y.}\ \bibnamefont
  {Wang}}, \bibinfo {author} {\bibfnamefont {C.}~\bibnamefont {Song}}, \bibinfo
  {author} {\bibfnamefont {B.}~\bibnamefont {Cui}}, \bibinfo {author}
  {\bibfnamefont {G.~Y.}\ \bibnamefont {Wang}}, \bibinfo {author}
  {\bibfnamefont {F.}~\bibnamefont {Zeng}}, \ and\ \bibinfo {author}
  {\bibfnamefont {F.}~\bibnamefont {Pan}},\ }\bibfield  {title} {\enquote
  {\bibinfo {title} {Room-temperature perpendicular exchange coupling and
  tunneling anisotropic magnetoresistance in an antiferromagnet-based tunnel
  junction},}\ }\href {\doibase 10.1103/PhysRevLett.109.137201} {\bibfield
  {journal} {\bibinfo  {journal} {Phys. Rev. Lett.}\ }\textbf {\bibinfo
  {volume} {109}},\ \bibinfo {pages} {137201} (\bibinfo {year}
  {2012})}\BibitemShut {NoStop}%
\bibitem [{\citenamefont {Yan}\ \emph {et~al.}(2019)\citenamefont {Yan},
  \citenamefont {Feng}, \citenamefont {Shang}, \citenamefont {Wang},
  \citenamefont {Hu}, \citenamefont {Wang}, \citenamefont {Zhu}, \citenamefont
  {Wang}, \citenamefont {Chen}, \citenamefont {Hua}, \citenamefont {Lu},
  \citenamefont {Wang}, \citenamefont {Qin}, \citenamefont {Guo}, \citenamefont
  {Zhou}, \citenamefont {Leng}, \citenamefont {Liu}, \citenamefont {Jiang},
  \citenamefont {Coey},\ and\ \citenamefont {Liu}}]{MnPt_Piezo_NNano19}%
  \BibitemOpen
  \bibfield  {author} {\bibinfo {author} {\bibfnamefont {Han}\ \bibnamefont
  {Yan}}, \bibinfo {author} {\bibfnamefont {Zexin}\ \bibnamefont {Feng}},
  \bibinfo {author} {\bibfnamefont {Shunli}\ \bibnamefont {Shang}}, \bibinfo
  {author} {\bibfnamefont {Xiaoning}\ \bibnamefont {Wang}}, \bibinfo {author}
  {\bibfnamefont {Zexiang}\ \bibnamefont {Hu}}, \bibinfo {author}
  {\bibfnamefont {Jinhua}\ \bibnamefont {Wang}}, \bibinfo {author}
  {\bibfnamefont {Zengwei}\ \bibnamefont {Zhu}}, \bibinfo {author}
  {\bibfnamefont {Hui}\ \bibnamefont {Wang}}, \bibinfo {author} {\bibfnamefont
  {Zuhuang}\ \bibnamefont {Chen}}, \bibinfo {author} {\bibfnamefont {Hui}\
  \bibnamefont {Hua}}, \bibinfo {author} {\bibfnamefont {Wenkuo}\ \bibnamefont
  {Lu}}, \bibinfo {author} {\bibfnamefont {Jingmin}\ \bibnamefont {Wang}},
  \bibinfo {author} {\bibfnamefont {Peixin}\ \bibnamefont {Qin}}, \bibinfo
  {author} {\bibfnamefont {Huixin}\ \bibnamefont {Guo}}, \bibinfo {author}
  {\bibfnamefont {Xiaorong}\ \bibnamefont {Zhou}}, \bibinfo {author}
  {\bibfnamefont {Zhaoguogang}\ \bibnamefont {Leng}}, \bibinfo {author}
  {\bibfnamefont {Zikui}\ \bibnamefont {Liu}}, \bibinfo {author} {\bibfnamefont
  {Chengbao}\ \bibnamefont {Jiang}}, \bibinfo {author} {\bibfnamefont
  {Michael}\ \bibnamefont {Coey}}, \ and\ \bibinfo {author} {\bibfnamefont
  {Zhiqi}\ \bibnamefont {Liu}},\ }\bibfield  {title} {\enquote {\bibinfo
  {title} {{A piezoelectric, strain-controlled antiferromagnetic memory
  insensitive to magnetic fields}},}\ }\href {\doibase
  10.1038/s41565-018-0339-0} {\bibfield  {journal} {\bibinfo  {journal} {Nature
  Nanotechnology}\ }\textbf {\bibinfo {volume} {14}},\ \bibinfo {pages}
  {131--136} (\bibinfo {year} {2019})}\BibitemShut {NoStop}%
\bibitem [{\citenamefont {Park}\ \emph {et~al.}(2019)\citenamefont {Park},
  \citenamefont {Lee}, \citenamefont {Das}, \citenamefont {Debnath},
  \citenamefont {Carman},\ and\ \citenamefont {Lake}}]{InJunPark_APL19}%
  \BibitemOpen
  \bibfield  {author} {\bibinfo {author} {\bibfnamefont {I.~J.}\ \bibnamefont
  {Park}}, \bibinfo {author} {\bibfnamefont {T.}~\bibnamefont {Lee}}, \bibinfo
  {author} {\bibfnamefont {P.}~\bibnamefont {Das}}, \bibinfo {author}
  {\bibfnamefont {B.}~\bibnamefont {Debnath}}, \bibinfo {author} {\bibfnamefont
  {G.~P.}\ \bibnamefont {Carman}}, \ and\ \bibinfo {author} {\bibfnamefont
  {R.~K.}\ \bibnamefont {Lake}},\ }\bibfield  {title} {\enquote {\bibinfo
  {title} {Strain control of the {N\'{e}el} vector in {Mn}-based
  antiferromagnets},}\ }\href {\doibase 10.1063/1.5093701} {\bibfield
  {journal} {\bibinfo  {journal} {Appl. Phys. Lett.}\ }\textbf {\bibinfo
  {volume} {114}},\ \bibinfo {pages} {142403} (\bibinfo {year}
  {2019})}\BibitemShut {NoStop}%
\bibitem [{\citenamefont {Velkov}\ \emph {et~al.}(2016)\citenamefont {Velkov},
  \citenamefont {Gomonay}, \citenamefont {Beens}, \citenamefont {Schwiete},
  \citenamefont {Brataas}, \citenamefont {Sinova},\ and\ \citenamefont
  {Duine}}]{2016_J_ind_AFM-Sky_motion_NJP}%
  \BibitemOpen
  \bibfield  {author} {\bibinfo {author} {\bibfnamefont {H}~\bibnamefont
  {Velkov}}, \bibinfo {author} {\bibfnamefont {O}~\bibnamefont {Gomonay}},
  \bibinfo {author} {\bibfnamefont {M}~\bibnamefont {Beens}}, \bibinfo {author}
  {\bibfnamefont {G}~\bibnamefont {Schwiete}}, \bibinfo {author} {\bibfnamefont
  {A}~\bibnamefont {Brataas}}, \bibinfo {author} {\bibfnamefont
  {J}~\bibnamefont {Sinova}}, \ and\ \bibinfo {author} {\bibfnamefont {R~A}\
  \bibnamefont {Duine}},\ }\bibfield  {title} {\enquote {\bibinfo {title}
  {Phenomenology of current-induced skyrmion motion in antiferromagnets},}\
  }\href {\doibase 10.1088/1367-2630/18/7/075016} {\bibfield  {journal}
  {\bibinfo  {journal} {New Journal of Physics}\ }\textbf {\bibinfo {volume}
  {18}},\ \bibinfo {pages} {075016} (\bibinfo {year} {2016})}\BibitemShut
  {NoStop}%
\bibitem [{\citenamefont {Jin}\ \emph {et~al.}(2016)\citenamefont {Jin},
  \citenamefont {Song}, \citenamefont {Wang},\ and\ \citenamefont
  {Liu}}]{AFM_driven_by_SHE_APL16}%
  \BibitemOpen
  \bibfield  {author} {\bibinfo {author} {\bibfnamefont {Chendong}\
  \bibnamefont {Jin}}, \bibinfo {author} {\bibfnamefont {Chengkun}\
  \bibnamefont {Song}}, \bibinfo {author} {\bibfnamefont {Jianbo}\ \bibnamefont
  {Wang}}, \ and\ \bibinfo {author} {\bibfnamefont {Qingfang}\ \bibnamefont
  {Liu}},\ }\bibfield  {title} {\enquote {\bibinfo {title} {Dynamics of
  antiferromagnetic skyrmion driven by the spin hall effect},}\ }\href
  {\doibase 10.1063/1.4967006} {\bibfield  {journal} {\bibinfo  {journal}
  {Applied Physics Letters}\ }\textbf {\bibinfo {volume} {109}},\ \bibinfo
  {pages} {182404} (\bibinfo {year} {2016})}\BibitemShut {NoStop}%
\bibitem [{\citenamefont {Zhang}\ \emph {et~al.}(2016)\citenamefont {Zhang},
  \citenamefont {Zhou},\ and\ \citenamefont
  {Ezawa}}]{2016_AFM_Skx_Stab_Creat_Manip_Ezawa_SciRep}%
  \BibitemOpen
  \bibfield  {author} {\bibinfo {author} {\bibfnamefont {Xichao}\ \bibnamefont
  {Zhang}}, \bibinfo {author} {\bibfnamefont {Yan}\ \bibnamefont {Zhou}}, \
  and\ \bibinfo {author} {\bibfnamefont {Motohiko}\ \bibnamefont {Ezawa}},\
  }\bibfield  {title} {\enquote {\bibinfo {title} {{Antiferromagnetic Skyrmion:
  Stability, Creation and Manipulation}},}\ }\href {doi.org/10.1038/srep24795}
  {\bibfield  {journal} {\bibinfo  {journal} {Scientific Reports}\ }\textbf
  {\bibinfo {volume} {6}},\ \bibinfo {pages} {24795} (\bibinfo {year}
  {2016})}\BibitemShut {NoStop}%
\bibitem [{\citenamefont {Barker}\ and\ \citenamefont
  {Tretiakov}(2016)}]{2016_AFM-Skx_J_T_Tretiakov_PRL}%
  \BibitemOpen
  \bibfield  {author} {\bibinfo {author} {\bibfnamefont {Joseph}\ \bibnamefont
  {Barker}}\ and\ \bibinfo {author} {\bibfnamefont {Oleg~A.}\ \bibnamefont
  {Tretiakov}},\ }\bibfield  {title} {\enquote {\bibinfo {title} {Static and
  dynamical properties of antiferromagnetic skyrmions in the presence of
  applied current and temperature},}\ }\href {\doibase
  10.1103/PhysRevLett.116.147203} {\bibfield  {journal} {\bibinfo  {journal}
  {Phys. Rev. Lett.}\ }\textbf {\bibinfo {volume} {116}},\ \bibinfo {pages}
  {147203} (\bibinfo {year} {2016})}\BibitemShut {NoStop}%
\bibitem [{\citenamefont {G\"obel}\ \emph {et~al.}(2017)\citenamefont
  {G\"obel}, \citenamefont {Mook}, \citenamefont {Henk},\ and\ \citenamefont
  {Mertig}}]{2017_AFM_Skx_crystals_PRB}%
  \BibitemOpen
  \bibfield  {author} {\bibinfo {author} {\bibfnamefont {B\"orge}\ \bibnamefont
  {G\"obel}}, \bibinfo {author} {\bibfnamefont {Alexander}\ \bibnamefont
  {Mook}}, \bibinfo {author} {\bibfnamefont {J\"urgen}\ \bibnamefont {Henk}}, \
  and\ \bibinfo {author} {\bibfnamefont {Ingrid}\ \bibnamefont {Mertig}},\
  }\bibfield  {title} {\enquote {\bibinfo {title} {Antiferromagnetic skyrmion
  crystals: Generation, topological hall, and topological spin hall effect},}\
  }\href {\doibase 10.1103/PhysRevB.96.060406} {\bibfield  {journal} {\bibinfo
  {journal} {Phys. Rev. B}\ }\textbf {\bibinfo {volume} {96}},\ \bibinfo
  {pages} {060406} (\bibinfo {year} {2017})}\BibitemShut {NoStop}%
\bibitem [{\citenamefont {Akosa}\ \emph {et~al.}(2018)\citenamefont {Akosa},
  \citenamefont {Tretiakov}, \citenamefont {Tatara},\ and\ \citenamefont
  {Manchon}}]{THE_AFM_J_induced_motion_Tretiakov_PRL18}%
  \BibitemOpen
  \bibfield  {author} {\bibinfo {author} {\bibfnamefont {C.~A.}\ \bibnamefont
  {Akosa}}, \bibinfo {author} {\bibfnamefont {O.~A.}\ \bibnamefont
  {Tretiakov}}, \bibinfo {author} {\bibfnamefont {G.}~\bibnamefont {Tatara}}, \
  and\ \bibinfo {author} {\bibfnamefont {A.}~\bibnamefont {Manchon}},\
  }\bibfield  {title} {\enquote {\bibinfo {title} {Theory of the topological
  spin hall effect in antiferromagnetic skyrmions: Impact on current-induced
  motion},}\ }\href {\doibase 10.1103/PhysRevLett.121.097204} {\bibfield
  {journal} {\bibinfo  {journal} {Phys. Rev. Lett.}\ }\textbf {\bibinfo
  {volume} {121}},\ \bibinfo {pages} {097204} (\bibinfo {year}
  {2018})}\BibitemShut {NoStop}%
\bibitem [{\citenamefont {Gan}\ \emph {et~al.}(2018)\citenamefont {Gan},
  \citenamefont {Krishnia},\ and\ \citenamefont
  {Lew}}]{2018_Skx_injection_NJP}%
  \BibitemOpen
  \bibfield  {author} {\bibinfo {author} {\bibfnamefont {W.~L.}\ \bibnamefont
  {Gan}}, \bibinfo {author} {\bibfnamefont {S.}~\bibnamefont {Krishnia}}, \
  and\ \bibinfo {author} {\bibfnamefont {W.~S.}\ \bibnamefont {Lew}},\
  }\bibfield  {title} {\enquote {\bibinfo {title} {Efficient in-line skyrmion
  injection method for synthetic antiferromagnetic systems},}\ }\href
  {http://stacks.iop.org/1367-2630/20/i=1/a=013029} {\bibfield  {journal}
  {\bibinfo  {journal} {New Journal of Physics}\ }\textbf {\bibinfo {volume}
  {20}},\ \bibinfo {pages} {013029} (\bibinfo {year} {2018})}\BibitemShut
  {NoStop}%
\bibitem [{\citenamefont {Zhao}\ \emph {et~al.}(2018)\citenamefont {Zhao},
  \citenamefont {Ren}, \citenamefont {Xie},\ and\ \citenamefont
  {Liu}}]{AFM_Skx_FET_APL18}%
  \BibitemOpen
  \bibfield  {author} {\bibinfo {author} {\bibfnamefont {X.}~\bibnamefont
  {Zhao}}, \bibinfo {author} {\bibfnamefont {R.}~\bibnamefont {Ren}}, \bibinfo
  {author} {\bibfnamefont {G.}~\bibnamefont {Xie}}, \ and\ \bibinfo {author}
  {\bibfnamefont {Y.}~\bibnamefont {Liu}},\ }\bibfield  {title} {\enquote
  {\bibinfo {title} {Single antiferromagnetic skyrmion transistor based on
  strain manipulation},}\ }\href {\doibase 10.1063/1.5034515} {\bibfield
  {journal} {\bibinfo  {journal} {Appl. Phys. Lett.}\ }\textbf {\bibinfo
  {volume} {112}},\ \bibinfo {pages} {252402} (\bibinfo {year}
  {2018})}\BibitemShut {NoStop}%
\bibitem [{\citenamefont {Bessarab}\ \emph {et~al.}(2019)\citenamefont
  {Bessarab}, \citenamefont {Yudin}, \citenamefont {Gulevich}, \citenamefont
  {Wadley}, \citenamefont {Titov},\ and\ \citenamefont
  {Tretiakov}}]{Stability_lifetime_Tretiakov_PRB19}%
  \BibitemOpen
  \bibfield  {author} {\bibinfo {author} {\bibfnamefont {P.~F.}\ \bibnamefont
  {Bessarab}}, \bibinfo {author} {\bibfnamefont {D.}~\bibnamefont {Yudin}},
  \bibinfo {author} {\bibfnamefont {D.~R.}\ \bibnamefont {Gulevich}}, \bibinfo
  {author} {\bibfnamefont {P.}~\bibnamefont {Wadley}}, \bibinfo {author}
  {\bibfnamefont {M.}~\bibnamefont {Titov}}, \ and\ \bibinfo {author}
  {\bibfnamefont {Oleg~A.}\ \bibnamefont {Tretiakov}},\ }\bibfield  {title}
  {\enquote {\bibinfo {title} {Stability and lifetime of antiferromagnetic
  skyrmions},}\ }\href {\doibase 10.1103/PhysRevB.99.140411} {\bibfield
  {journal} {\bibinfo  {journal} {Phys. Rev. B}\ }\textbf {\bibinfo {volume}
  {99}},\ \bibinfo {pages} {140411} (\bibinfo {year} {2019})}\BibitemShut
  {NoStop}%
\bibitem [{\citenamefont {Khoshlahni}\ \emph {et~al.}(2019)\citenamefont
  {Khoshlahni}, \citenamefont {Qaiumzadeh}, \citenamefont {Bergman},\ and\
  \citenamefont {Brataas}}]{2019_UFast_Gen_Dyn_Brataas_PRB}%
  \BibitemOpen
  \bibfield  {author} {\bibinfo {author} {\bibfnamefont {Rohollah}\
  \bibnamefont {Khoshlahni}}, \bibinfo {author} {\bibfnamefont {Alireza}\
  \bibnamefont {Qaiumzadeh}}, \bibinfo {author} {\bibfnamefont {Anders}\
  \bibnamefont {Bergman}}, \ and\ \bibinfo {author} {\bibfnamefont {Arne}\
  \bibnamefont {Brataas}},\ }\bibfield  {title} {\enquote {\bibinfo {title}
  {Ultrafast generation and dynamics of isolated skyrmions in antiferromagnetic
  insulators},}\ }\href {\doibase 10.1103/PhysRevB.99.054423} {\bibfield
  {journal} {\bibinfo  {journal} {Phys. Rev. B}\ }\textbf {\bibinfo {volume}
  {99}},\ \bibinfo {pages} {054423} (\bibinfo {year} {2019})}\BibitemShut
  {NoStop}%
\bibitem [{\citenamefont {Legrand}\ \emph
  {et~al.}(2020{\natexlab{a}})\citenamefont {Legrand}, \citenamefont
  {Maccariello}, \citenamefont {Ajejas}, \citenamefont {Collin}, \citenamefont
  {Vecchiola}, \citenamefont {Bouzehouane}, \citenamefont {Reyren},
  \citenamefont {Cros},\ and\ \citenamefont
  {Fert}}]{RT_Stabilization_ASKn_Fert19}%
  \BibitemOpen
  \bibfield  {author} {\bibinfo {author} {\bibfnamefont {William}\ \bibnamefont
  {Legrand}}, \bibinfo {author} {\bibfnamefont {Davide}\ \bibnamefont
  {Maccariello}}, \bibinfo {author} {\bibfnamefont {Fernando}\ \bibnamefont
  {Ajejas}}, \bibinfo {author} {\bibfnamefont {Sophie}\ \bibnamefont {Collin}},
  \bibinfo {author} {\bibfnamefont {Aymeric}\ \bibnamefont {Vecchiola}},
  \bibinfo {author} {\bibfnamefont {Karim}\ \bibnamefont {Bouzehouane}},
  \bibinfo {author} {\bibfnamefont {Nicolas}\ \bibnamefont {Reyren}}, \bibinfo
  {author} {\bibfnamefont {Vincent}\ \bibnamefont {Cros}}, \ and\ \bibinfo
  {author} {\bibfnamefont {Albert}\ \bibnamefont {Fert}},\ }\bibfield  {title}
  {\enquote {\bibinfo {title} {{Room-temperature stabilization of
  antiferromagnetic skyrmions in synthetic antiferromagnets}},}\ }\href
  {\doibase 10.1038/s41563-019-0468-3} {\bibfield  {journal} {\bibinfo
  {journal} {Nature Materials}\ }\textbf {\bibinfo {volume} {19}},\ \bibinfo
  {pages} {34--42} (\bibinfo {year} {2020}{\natexlab{a}})}\BibitemShut
  {NoStop}%
\bibitem [{\citenamefont {Saha}\ \emph {et~al.}(2019)\citenamefont {Saha},
  \citenamefont {Srivastava}, \citenamefont {Ma}, \citenamefont {Jena},
  \citenamefont {Werner}, \citenamefont {Kumar}, \citenamefont {Felser},\ and\
  \citenamefont {Parkin}}]{Felser_Parkin_ASk_Heusler_NCom19}%
  \BibitemOpen
  \bibfield  {author} {\bibinfo {author} {\bibfnamefont {Rana}\ \bibnamefont
  {Saha}}, \bibinfo {author} {\bibfnamefont {Abhay~K.}\ \bibnamefont
  {Srivastava}}, \bibinfo {author} {\bibfnamefont {Tianping}\ \bibnamefont
  {Ma}}, \bibinfo {author} {\bibfnamefont {Jagannath}\ \bibnamefont {Jena}},
  \bibinfo {author} {\bibfnamefont {Peter}\ \bibnamefont {Werner}}, \bibinfo
  {author} {\bibfnamefont {Vivek}\ \bibnamefont {Kumar}}, \bibinfo {author}
  {\bibfnamefont {Claudia}\ \bibnamefont {Felser}}, \ and\ \bibinfo {author}
  {\bibfnamefont {Stuart S.~P.}\ \bibnamefont {Parkin}},\ }\bibfield  {title}
  {\enquote {\bibinfo {title} {Intrinsic stability of magnetic anti-skyrmions
  in the tetragonal inverse heusler compound
  {Mn$_{1.4}$Pt$_{0.9}$Pd$_{0.1}$Sn}},}\ }\href {\doibase
  10.1038/s41467-019-13323-x} {\bibfield  {journal} {\bibinfo  {journal}
  {Nature Communications}\ }\textbf {\bibinfo {volume} {10}},\ \bibinfo {pages}
  {5305} (\bibinfo {year} {2019})}\BibitemShut {NoStop}%
\bibitem [{\citenamefont {Buhl}\ \emph {et~al.}(2017)\citenamefont {Buhl},
  \citenamefont {Freimuth}, \citenamefont {Bl{\"{u}}gel},\ and\ \citenamefont
  {Mokrousov}}]{Buhl2017}%
  \BibitemOpen
  \bibfield  {author} {\bibinfo {author} {\bibfnamefont {Patrick~M.}\
  \bibnamefont {Buhl}}, \bibinfo {author} {\bibfnamefont {Frank}\ \bibnamefont
  {Freimuth}}, \bibinfo {author} {\bibfnamefont {Stefan}\ \bibnamefont
  {Bl{\"{u}}gel}}, \ and\ \bibinfo {author} {\bibfnamefont {Yuriy}\
  \bibnamefont {Mokrousov}},\ }\bibfield  {title} {\enquote {\bibinfo {title}
  {{Topological spin Hall effect in antiferromagnetic skyrmions}},}\ }\href
  {\doibase 10.1002/pssr.201700007} {\bibfield  {journal} {\bibinfo  {journal}
  {Physica Status Solidi - Rapid Research Letters}\ }\textbf {\bibinfo {volume}
  {11}},\ \bibinfo {pages} {1700007} (\bibinfo {year} {2017})}\BibitemShut
  {NoStop}%
\bibitem [{\citenamefont {Legrand}\ \emph
  {et~al.}(2020{\natexlab{b}})\citenamefont {Legrand}, \citenamefont
  {Maccariello}, \citenamefont {Ajejas}, \citenamefont {Collin}, \citenamefont
  {Vecchiola}, \citenamefont {Bouzehouane}, \citenamefont {Reyren},
  \citenamefont {Cros},\ and\ \citenamefont {Fert}}]{Legrand2019}%
  \BibitemOpen
  \bibfield  {author} {\bibinfo {author} {\bibfnamefont {William}\ \bibnamefont
  {Legrand}}, \bibinfo {author} {\bibfnamefont {Davide}\ \bibnamefont
  {Maccariello}}, \bibinfo {author} {\bibfnamefont {Fernando}\ \bibnamefont
  {Ajejas}}, \bibinfo {author} {\bibfnamefont {Sophie}\ \bibnamefont {Collin}},
  \bibinfo {author} {\bibfnamefont {Aymeric}\ \bibnamefont {Vecchiola}},
  \bibinfo {author} {\bibfnamefont {Karim}\ \bibnamefont {Bouzehouane}},
  \bibinfo {author} {\bibfnamefont {Nicolas}\ \bibnamefont {Reyren}}, \bibinfo
  {author} {\bibfnamefont {Vincent}\ \bibnamefont {Cros}}, \ and\ \bibinfo
  {author} {\bibfnamefont {Albert}\ \bibnamefont {Fert}},\ }\bibfield  {title}
  {\enquote {\bibinfo {title} {{Room-temperature stabilization of
  antiferromagnetic skyrmions in synthetic antiferromagnets}},}\ }\href
  {\doibase 10.1038/s41563-019-0468-3} {\bibfield  {journal} {\bibinfo
  {journal} {Nature Materials}\ }\textbf {\bibinfo {volume} {19}},\ \bibinfo
  {pages} {34--42} (\bibinfo {year} {2020}{\natexlab{b}})}\BibitemShut
  {NoStop}%
\bibitem [{\citenamefont {Ikegawa}\ \emph {et~al.}(2020)\citenamefont
  {Ikegawa}, \citenamefont {Mancoff}, \citenamefont {Janesky},\ and\
  \citenamefont {Aggarwal}}]{MRAM_Review_TED20}%
  \BibitemOpen
  \bibfield  {author} {\bibinfo {author} {\bibfnamefont {S.}~\bibnamefont
  {Ikegawa}}, \bibinfo {author} {\bibfnamefont {F.~B.}\ \bibnamefont
  {Mancoff}}, \bibinfo {author} {\bibfnamefont {J.}~\bibnamefont {Janesky}}, \
  and\ \bibinfo {author} {\bibfnamefont {S.}~\bibnamefont {Aggarwal}},\
  }\bibfield  {title} {\enquote {\bibinfo {title} {Magnetoresistive random
  access memory: Present and future},}\ }\href {\doibase
  10.1109/TED.2020.2965403} {\bibfield  {journal} {\bibinfo  {journal} {IEEE
  Trans. Elect. Dev.}\ }\textbf {\bibinfo {volume} {67}},\ \bibinfo {pages}
  {1407 -- 1418} (\bibinfo {year} {2020})}\BibitemShut {NoStop}%
\bibitem [{\citenamefont {Hanneken}\ \emph {et~al.}(2015)\citenamefont
  {Hanneken}, \citenamefont {Otte}, \citenamefont {Kubetzka}, \citenamefont
  {Dup{\'{e}}}, \citenamefont {Romming}, \citenamefont {von Bergmann},
  \citenamefont {Wiesendanger},\ and\ \citenamefont {Heinze}}]{Hanneken2015a}%
  \BibitemOpen
  \bibfield  {author} {\bibinfo {author} {\bibfnamefont {Christian}\
  \bibnamefont {Hanneken}}, \bibinfo {author} {\bibfnamefont {Fabian}\
  \bibnamefont {Otte}}, \bibinfo {author} {\bibfnamefont {Andr{\'{e}}}\
  \bibnamefont {Kubetzka}}, \bibinfo {author} {\bibfnamefont {Bertrand}\
  \bibnamefont {Dup{\'{e}}}}, \bibinfo {author} {\bibfnamefont {Niklas}\
  \bibnamefont {Romming}}, \bibinfo {author} {\bibfnamefont {Kirsten}\
  \bibnamefont {von Bergmann}}, \bibinfo {author} {\bibfnamefont {Roland}\
  \bibnamefont {Wiesendanger}}, \ and\ \bibinfo {author} {\bibfnamefont
  {Stefan}\ \bibnamefont {Heinze}},\ }\bibfield  {title} {\enquote {\bibinfo
  {title} {{Electrical detection of magnetic skyrmions by tunnelling
  non-collinear magnetoresistance}},}\ }\href {\doibase 10.1038/nnano.2015.218}
  {\bibfield  {journal} {\bibinfo  {journal} {Nature Nanotechnology}\ }\textbf
  {\bibinfo {volume} {10}},\ \bibinfo {pages} {1039--1042} (\bibinfo {year}
  {2015})}\BibitemShut {NoStop}%
\bibitem [{\citenamefont {Kubetzka}\ \emph {et~al.}(2017)\citenamefont
  {Kubetzka}, \citenamefont {Hanneken}, \citenamefont {Wiesendanger},\ and\
  \citenamefont {{Von Bergmann}}}]{Kubetzka2017}%
  \BibitemOpen
  \bibfield  {author} {\bibinfo {author} {\bibfnamefont {Andr{\'{e}}}\
  \bibnamefont {Kubetzka}}, \bibinfo {author} {\bibfnamefont {Christian}\
  \bibnamefont {Hanneken}}, \bibinfo {author} {\bibfnamefont {Roland}\
  \bibnamefont {Wiesendanger}}, \ and\ \bibinfo {author} {\bibfnamefont
  {Kirsten}\ \bibnamefont {{Von Bergmann}}},\ }\bibfield  {title} {\enquote
  {\bibinfo {title} {{Impact of the skyrmion spin texture on
  magnetoresistance}},}\ }\href {\doibase 10.1103/PhysRevB.95.104433}
  {\bibfield  {journal} {\bibinfo  {journal} {Physical Review B}\ }\textbf
  {\bibinfo {volume} {95}},\ \bibinfo {pages} {104433} (\bibinfo {year}
  {2017})},\ \Eprint {http://arxiv.org/abs/1701.09077} {arXiv:1701.09077}
  \BibitemShut {NoStop}%
\bibitem [{\citenamefont {Maccariello}\ \emph {et~al.}(2018)\citenamefont
  {Maccariello}, \citenamefont {Legrand}, \citenamefont {Reyren}, \citenamefont
  {Garcia}, \citenamefont {Bouzehouane}, \citenamefont {Collin}, \citenamefont
  {Cros},\ and\ \citenamefont {Fert}}]{Maccariello2018a}%
  \BibitemOpen
  \bibfield  {author} {\bibinfo {author} {\bibfnamefont {Davide}\ \bibnamefont
  {Maccariello}}, \bibinfo {author} {\bibfnamefont {William}\ \bibnamefont
  {Legrand}}, \bibinfo {author} {\bibfnamefont {Nicolas}\ \bibnamefont
  {Reyren}}, \bibinfo {author} {\bibfnamefont {Karin}\ \bibnamefont {Garcia}},
  \bibinfo {author} {\bibfnamefont {Karim}\ \bibnamefont {Bouzehouane}},
  \bibinfo {author} {\bibfnamefont {Sophie}\ \bibnamefont {Collin}}, \bibinfo
  {author} {\bibfnamefont {Vincent}\ \bibnamefont {Cros}}, \ and\ \bibinfo
  {author} {\bibfnamefont {Albert}\ \bibnamefont {Fert}},\ }\bibfield  {title}
  {\enquote {\bibinfo {title} {{Electrical detection of single magnetic
  skyrmions in metallic multilayers at room temperature}},}\ }\href {\doibase
  10.1038/s41565-017-0044-4} {\bibfield  {journal} {\bibinfo  {journal} {Nature
  Nanotechnology}\ }\textbf {\bibinfo {volume} {13}},\ \bibinfo {pages}
  {233--237} (\bibinfo {year} {2018})}\BibitemShut {NoStop}%
\bibitem [{\citenamefont {Wang}\ \emph {et~al.}(2019)\citenamefont {Wang},
  \citenamefont {Tang}, \citenamefont {Wang}, \citenamefont {Kong},
  \citenamefont {Tian},\ and\ \citenamefont {Du}}]{Wang2019}%
  \BibitemOpen
  \bibfield  {author} {\bibinfo {author} {\bibfnamefont {Shasha}\ \bibnamefont
  {Wang}}, \bibinfo {author} {\bibfnamefont {Jin}\ \bibnamefont {Tang}},
  \bibinfo {author} {\bibfnamefont {Weiwei}\ \bibnamefont {Wang}}, \bibinfo
  {author} {\bibfnamefont {Lingyao}\ \bibnamefont {Kong}}, \bibinfo {author}
  {\bibfnamefont {Mingliang}\ \bibnamefont {Tian}}, \ and\ \bibinfo {author}
  {\bibfnamefont {Haifeng}\ \bibnamefont {Du}},\ }\bibfield  {title} {\enquote
  {\bibinfo {title} {{Electrical Detection of Magnetic Skyrmions}},}\ }\href
  {\doibase 10.1007/s10909-019-02202-w} {\bibfield  {journal} {\bibinfo
  {journal} {Journal of Low Temperature Physics}\ }\textbf {\bibinfo {volume}
  {197}},\ \bibinfo {pages} {321--336} (\bibinfo {year} {2019})}\BibitemShut
  {NoStop}%
\bibitem [{\citenamefont {Hamamoto}\ \emph {et~al.}(2016)\citenamefont
  {Hamamoto}, \citenamefont {Ezawa},\ and\ \citenamefont
  {Nagaosa}}]{Hamamoto2016}%
  \BibitemOpen
  \bibfield  {author} {\bibinfo {author} {\bibfnamefont {Keita}\ \bibnamefont
  {Hamamoto}}, \bibinfo {author} {\bibfnamefont {Motohiko}\ \bibnamefont
  {Ezawa}}, \ and\ \bibinfo {author} {\bibfnamefont {Naoto}\ \bibnamefont
  {Nagaosa}},\ }\bibfield  {title} {\enquote {\bibinfo {title} {{Purely
  electrical detection of a skyrmion in constricted geometry}},}\ }\href
  {\doibase 10.1063/1.4943949} {\bibfield  {journal} {\bibinfo  {journal}
  {Applied Physics Letters}\ }\textbf {\bibinfo {volume} {108}},\ \bibinfo
  {pages} {112401} (\bibinfo {year} {2016})}\BibitemShut {NoStop}%
\bibitem [{\citenamefont {Tomasello}\ \emph {et~al.}(2017)\citenamefont
  {Tomasello}, \citenamefont {Ricci}, \citenamefont {Burrascano}, \citenamefont
  {Puliafito}, \citenamefont {Carpentieri},\ and\ \citenamefont
  {Finocchio}}]{Tomasello2017}%
  \BibitemOpen
  \bibfield  {author} {\bibinfo {author} {\bibfnamefont {Riccardo}\
  \bibnamefont {Tomasello}}, \bibinfo {author} {\bibfnamefont {Marco}\
  \bibnamefont {Ricci}}, \bibinfo {author} {\bibfnamefont {Pietro}\
  \bibnamefont {Burrascano}}, \bibinfo {author} {\bibfnamefont {Vito}\
  \bibnamefont {Puliafito}}, \bibinfo {author} {\bibfnamefont {Mario}\
  \bibnamefont {Carpentieri}}, \ and\ \bibinfo {author} {\bibfnamefont
  {Giovanni}\ \bibnamefont {Finocchio}},\ }\bibfield  {title} {\enquote
  {\bibinfo {title} {{Electrical detection of single magnetic skyrmion at room
  temperature}},}\ }\href {\doibase 10.1063/1.4975998} {\bibfield  {journal}
  {\bibinfo  {journal} {AIP Advances}\ }\textbf {\bibinfo {volume} {7}},\
  \bibinfo {pages} {056022} (\bibinfo {year} {2017})}\BibitemShut {NoStop}%
\bibitem [{\citenamefont {Stolt}\ \emph {et~al.}(2019)\citenamefont {Stolt},
  \citenamefont {Schneider}, \citenamefont {Mathur}, \citenamefont {Shearer},
  \citenamefont {Rellinghaus}, \citenamefont {Nielsch},\ and\ \citenamefont
  {Jin}}]{Stolt2019}%
  \BibitemOpen
  \bibfield  {author} {\bibinfo {author} {\bibfnamefont {Matthew~J.}\
  \bibnamefont {Stolt}}, \bibinfo {author} {\bibfnamefont {Sebastian}\
  \bibnamefont {Schneider}}, \bibinfo {author} {\bibfnamefont {Nitish}\
  \bibnamefont {Mathur}}, \bibinfo {author} {\bibfnamefont {Melinda~J.}\
  \bibnamefont {Shearer}}, \bibinfo {author} {\bibfnamefont {Bernd}\
  \bibnamefont {Rellinghaus}}, \bibinfo {author} {\bibfnamefont {Kornelius}\
  \bibnamefont {Nielsch}}, \ and\ \bibinfo {author} {\bibfnamefont {Song}\
  \bibnamefont {Jin}},\ }\bibfield  {title} {\enquote {\bibinfo {title}
  {Electrical detection and magnetic imaging of stabilized magnetic skyrmions
  in {Fe$_{1−x}$Co$_x$Ge} (x < 0.1) microplates},}\ }\href {\doibase
  10.1002/adfm.201805418} {\bibfield  {journal} {\bibinfo  {journal} {Advanced
  Functional Materials}\ }\textbf {\bibinfo {volume} {29}},\ \bibinfo {pages}
  {1805418} (\bibinfo {year} {2019})}\BibitemShut {NoStop}%
\bibitem [{\citenamefont {Sch\"afer-Richarz}\ \emph {et~al.}(2019)\citenamefont
  {Sch\"afer-Richarz}, \citenamefont {Risius}, \citenamefont {Czerner},\ and\
  \citenamefont {Heiliger}}]{MTJ_Skx_Detection_PRB19}%
  \BibitemOpen
  \bibfield  {author} {\bibinfo {author} {\bibfnamefont {Jonas~Friedrich}\
  \bibnamefont {Sch\"afer-Richarz}}, \bibinfo {author} {\bibfnamefont
  {Philipp}\ \bibnamefont {Risius}}, \bibinfo {author} {\bibfnamefont
  {Michael}\ \bibnamefont {Czerner}}, \ and\ \bibinfo {author} {\bibfnamefont
  {Christian}\ \bibnamefont {Heiliger}},\ }\bibfield  {title} {\enquote
  {\bibinfo {title} {Magnetic tunnel junctions: An efficient way for electrical
  skyrmion detection investigated by ab initio theory},}\ }\href {\doibase
  10.1103/PhysRevB.100.214413} {\bibfield  {journal} {\bibinfo  {journal}
  {Phys. Rev. B}\ }\textbf {\bibinfo {volume} {100}},\ \bibinfo {pages}
  {214413} (\bibinfo {year} {2019})}\BibitemShut {NoStop}%
\bibitem [{\citenamefont {Ohgushi}\ \emph {et~al.}(2000)\citenamefont
  {Ohgushi}, \citenamefont {Murakami},\ and\ \citenamefont
  {Nagaosa}}]{Nagaosa_Hunds_Rule_Coupling_PRL00}%
  \BibitemOpen
  \bibfield  {author} {\bibinfo {author} {\bibfnamefont {Kenya}\ \bibnamefont
  {Ohgushi}}, \bibinfo {author} {\bibfnamefont {Shuichi}\ \bibnamefont
  {Murakami}}, \ and\ \bibinfo {author} {\bibfnamefont {Naoto}\ \bibnamefont
  {Nagaosa}},\ }\bibfield  {title} {\enquote {\bibinfo {title} {Spin anisotropy
  and quantum hall effect in the kagom\'e lattice: Chiral spin state based on a
  ferromagnet},}\ }\href {\doibase 10.1103/PhysRevB.62.R6065} {\bibfield
  {journal} {\bibinfo  {journal} {Phys. Rev. B}\ }\textbf {\bibinfo {volume}
  {62}},\ \bibinfo {pages} {R6065--R6068} (\bibinfo {year} {2000})}\BibitemShut
  {NoStop}%
\bibitem [{\citenamefont {Sancho}\ \emph {et~al.}(1985)\citenamefont {Sancho},
  \citenamefont {Sancho},\ and\ \citenamefont
  {Rubio}}]{MPLSancho_HighlyConvergent_JPF85}%
  \BibitemOpen
  \bibfield  {author} {\bibinfo {author} {\bibfnamefont {M.~P.~Lopez}\
  \bibnamefont {Sancho}}, \bibinfo {author} {\bibfnamefont {J.~M.~Lopez}\
  \bibnamefont {Sancho}}, \ and\ \bibinfo {author} {\bibfnamefont
  {J.}~\bibnamefont {Rubio}},\ }\bibfield  {title} {\enquote {\bibinfo {title}
  {Highly convergent schemes for the calculation of bulk and surface green
  functions},}\ }\href@noop {} {\bibfield  {journal} {\bibinfo  {journal} {J.
  Phys. F}\ }\textbf {\bibinfo {volume} {15}},\ \bibinfo {pages} {851--858}
  (\bibinfo {year} {1985})}\BibitemShut {NoStop}%
\bibitem [{\citenamefont {Fano}(1961)}]{Ugo_Fano_PRL61}%
  \BibitemOpen
  \bibfield  {author} {\bibinfo {author} {\bibfnamefont {U.}~\bibnamefont
  {Fano}},\ }\bibfield  {title} {\enquote {\bibinfo {title} {Effects of
  configuration interaction on intensities and phase shifts},}\ }\href
  {\doibase 10.1103/PhysRev.124.1866} {\bibfield  {journal} {\bibinfo
  {journal} {Phys. Rev.}\ }\textbf {\bibinfo {volume} {124}},\ \bibinfo {pages}
  {1866--1878} (\bibinfo {year} {1961})}\BibitemShut {NoStop}%
\bibitem [{\citenamefont {Bowen}\ \emph {et~al.}(2005)\citenamefont {Bowen},
  \citenamefont {Barth{\'{e}}l{\'{e}}my}, \citenamefont {Bibes}, \citenamefont
  {Jacquet}, \citenamefont {Contour}, \citenamefont {Fert}, \citenamefont
  {Wortmann},\ and\ \citenamefont {Bl{\"{u}}gel}}]{Bowen2005}%
  \BibitemOpen
  \bibfield  {author} {\bibinfo {author} {\bibfnamefont {M.}~\bibnamefont
  {Bowen}}, \bibinfo {author} {\bibfnamefont {A.}~\bibnamefont
  {Barth{\'{e}}l{\'{e}}my}}, \bibinfo {author} {\bibfnamefont {M.}~\bibnamefont
  {Bibes}}, \bibinfo {author} {\bibfnamefont {E.}~\bibnamefont {Jacquet}},
  \bibinfo {author} {\bibfnamefont {J.~P.}\ \bibnamefont {Contour}}, \bibinfo
  {author} {\bibfnamefont {A.}~\bibnamefont {Fert}}, \bibinfo {author}
  {\bibfnamefont {D.}~\bibnamefont {Wortmann}}, \ and\ \bibinfo {author}
  {\bibfnamefont {S.}~\bibnamefont {Bl{\"{u}}gel}},\ }\bibfield  {title}
  {\enquote {\bibinfo {title} {{Half-metallicity proven using fully
  spin-polarized tunnelling}},}\ }\href {\doibase 10.1088/0953-8984/17/41/L02}
  {\bibfield  {journal} {\bibinfo  {journal} {Journal of Physics Condensed
  Matter}\ }\textbf {\bibinfo {volume} {17}},\ \bibinfo {pages} {L407--L409}
  (\bibinfo {year} {2005})}\BibitemShut {NoStop}%
\end{thebibliography}%

\end{document}